\documentclass[fleqn,usenatbib]{mnras}

\usepackage{newtxtext,newtxmath}
\usepackage[T1]{fontenc}

\DeclareRobustCommand{\VAN}[3]{#2}
\let\VANthebibliography\thebibliography
\def\thebibliography{\DeclareRobustCommand{\VAN}[3]{##3}\VANthebibliography}

\usepackage{graphicx}	% Including figure files
\usepackage{amsmath}	% Advanced maths commands
\title[Striation in Thin Channel Maps]{Nature of Striation in 21 cm Channel Maps: Velocity Caustics}

\author[Hu et al.]{
	Yue Hu$^{1,2}$\thanks{E-mail: yue.hu@wisc.edu}
	, A. Lazarian$^{2,3}$\thanks{E-mail: alazarian@facstaff.wisc.edu}, D. Alina$^{4}$, D. Pogosyan$^{5}$, Ka Wai Ho$^{2}$
	\\
	$^{1}$Department of Physics, University of Wisconsin-Madison, Madison, WI, 53706, USA\\
	$^{2}$Department of Astronomy, University of Wisconsin-Madison, Madison, WI, 53706, USA\\
	$^{3}$Centro de Investigación en Astronomía, Universidad Bernardo O’Higgins, Santiago, General Gana 1760, 8370993,
	Chile\\
	$^{4}$Department of Physics, School of Sciences and Humanities, Nazarbayev University, Astana 010000, Kazakhstan\\
	$^{5}$Department of Physics, University of Alberta, Edmonton, Alberta, T6G 2E1, Canada\\
}
\date{Accepted XXX. Received YYY; in original form ZZZ}

\pubyear{2023}

\begin{document}
	\label{firstpage}
	\pagerange{\pageref{firstpage}--\pageref{lastpage}}
	\maketitle
	
	% Abstract of the paper
	\begin{abstract}
		The alignment of striated intensity structures in thin neutral hydrogen (HI) spectroscopic channels with Galactic magnetic fields has been observed. However, the origin and nature of these striations are still debatable. Some studies suggest that the striations result solely from real cold-density filaments without considering the role of turbulent velocity fields in shaping the channel's intensity distribution. To determine the relative contribution of density and velocity in forming the striations in channel maps, we analyze synthetic observations of channel maps obtained from realistic magnetized multi-phase HI simulations with thermal broadening included. We vary the thickness of the channel maps and apply the Velocity Decomposition Algorithm to separate the velocity and density contributions. In parallel, we analyze GALFA-HI observations and compare the results. Our analysis shows that the thin channels are dominated by velocity contribution, and velocity caustics mainly generate the HI striations. We show that velocity caustics can cause a correlation between unsharp-masked HI structures and far-infrared emission. We demonstrate that the linear HI fibers revealed by the Rolling Hough Transform (RHT) in thin velocity channels originate from velocity caustics. As the thickness of channel maps increases, the relative contribution of density fluctuations in channel maps increases and more RHT-detected fibers tend to be perpendicular to the magnetic field. Conversely, the alignment with the magnetic field is the most prominent in thin channels. We conclude that similar to the Velocity Channel Gradients (VChGs) approach, RHT traces magnetic fields through the analysis of velocity caustics in thin channel maps.
	\end{abstract}

	% Select between one and six entries from the list of approved keywords.
	% Don't make up new ones.

	%%%%%%%%%%%%%%%%%%%%%%%%%%%%%%%%%%%%%%%%%%%%%%%%%%
	
	%%%%%%%%%%%%%%%%% BODY OF PAPER %%%%%%%%%%%%%%%%%%

\section{Introduction}
Neutral hydrogen (HI) is the most abundant element in the universe, providing crucial information about galaxies' structure and evolution \citep{1987ARA&A..25..303C,2016A&A...594A.116H,2018ApJS..234....2P}. In our Galaxy, studying the physical nature of HI  intensity structures is essential to understanding the interstellar medium (ISM; \citealt{1990ARA&A..28..215D}), star formation processes \citep{MK04, MO07, Crutcher12}, chemical evolution \citep{1989ApJ...338...48H,1997ARA&A..35..217W,1999RvMP...71..173H}, and Galactic dynamics \citep{1981gask.book.....M,1988gera.book...95K,2004ARA&A..42..211E}. 

The study of HI in astrophysics typically employs Position-Position-Velocity (PPV) cubes, where the sky coordinates of Position-Position are complemented with Doppler shift information in Velocity. The HI observation is commonly analyzed by slicing PPV cubes to create channel maps. Studies by \cite{1993MNRAS.262..327G} and \cite{1999MNRAS.302..417S} have shown that channel maps of the Milky Way and Small Magellanic Cloud respectively exhibit power-law statistics. The observed power spectrum and its variation with the thickness of the channel maps were interpreted as a consequence of HI's non-linear spectroscopic mapping from Position-Position-Position (PPP) space to PPV space \citep{LP00}. In the process of this mapping, HI clouds with different line-of-sight (LOS) positions but similar LOS velocities can be sampled into the same location in PPV space. This causes the HI's intensity distribution in PPV space to appear crowded, morphologically distorted, and statistically modified. This phenomenon, called velocity crowding, is significant in ISM that is known to be turbulent \citep{1995ApJ...443..209A, 2010ApJ...710..853C,2017ApJ...835....2X,2022ApJ...934....7H,2022arXiv220413760Y}. 

\cite{LP00} developed the theory describing the velocity crowding effect and demonstrated that its significance is linked to the thickness of the channel map. In thin channels (i.e., with high-velocity resolution), the observed intensity structures of HI are significantly distorted. These intensity distributions are primarily created by velocity crowding, so their statistics are controlled by velocity rather than density statistics. The subsequent numerical studies \citep{2003MNRAS.342..325E,2009ApJ...693.1074C,2009ApJ...707L.153P} confirmed the theoretical predictions. On the other side, it is well known that magnetohydrodynamic (MHD) turbulence is anisotropic, with turbulent eddies elongated along the magnetic fields \citep{GS95, LV99,2016JPlPh..82f5301F,2018MNRAS.481.5275T,2019tuma.book.....B,2020MNRAS.492..668B}. This anisotropy is imprinted in the velocity crowding so that the intensity structures in thin channels are striated along the magnetic fields, denoted as HI striations. The theory of this effect was elaborated in \cite{2016MNRAS.461.1227K} with anisotropies induced by three major MHD modes, solenoidal Alfv\'en mode, and compressive slow and fast modes, quantified (see also \citealt{2023MNRAS.521..530Y}).

The HI striations are critical for numerous studies, including tracing the Galactic magnetic field orientation at large scales \citep{YL17a, LY18a, HYL20, 2020MNRAS.496.2868L}, mapping the magnetic field strength of the diffuse ISM \citep{Lazarian18, 2023arXiv230205047H}, and modeling the Galactic foreground polarization \citep{Clark15, 2019ApJ...887..136C}. For instance, \cite{2002ASPC..276..182L} proposed to use the anisotropy of HI striations measured in thin channel maps to trace magnetic field directions and study magnetization, introducing a  new technique that was termed Velocity Channel Gradients (VChGs; \citealt{LY18a}). As an earlier separate development, \cite{2014ApJ...789...82C} used the linear-structure-detection algorithm, the Rolling Hough Transform (RHT), to identify the HI striations and reported an alignment with the Galactic magnetic field. However, the striations detected with RHT (denoted as RHT-fiber) were assumed to result from the actual Galactic HI cold density filaments \citep{Clark19}. 

The nature of striations in HI channel maps remains a hotly debated subject \citep{ Clark19, Yuen19, 2019ApJ...886L..13P, 2020arXiv200301454K,VDA,2023arXiv230316183K}. The controversy involves the applicability of the theory of \cite{LP00} to multiphase HI, where both cold and warm phases are present \citep{1977ApJ...218..148M}. In particular, according to \cite{Clark19}, the alignment of RHT-fibers in thin channel maps results from the alignment of cold density filaments with magnetic field, while the effect of velocity caustics is negligible. The overlapping correlation of unsharp-masked (see \S~\ref{sec:dis}) GALFA-HI data \citep{2018ApJS..234....2P} and far-infrared (FIR) Planck observations \citep{2020A&A...641A...3P} was employed to support this interpretation in \cite{Clark19}.  However, the statistical significance of the correlation between HI and Planck was assessed using an un-normalized parameter. It was shown in \cite{Yuen19} that after a proper normalization was introduced, this correlation is insignificant (see also Appendix~\ref{app: I857}). In contrast, according to \cite{LY18a,Yuen19,HLY20,2020MNRAS.496.2868L,VDA}, the striation revealed with the VChGS naturally arises from the anisotropy of velocity fluctuations in MHD turbulence. 

In the present paper, we attempt to resolve the controversy above by exploring two approaches. The observational one separates channel maps' velocity and density contributions and evaluates their relative significance. This approach has already been explored in \citep{VDA} where the Velocity Decomposition Algorithm (VDA; \citealt{VDA}) was introduced and successfully applied to GALFA-HI data. The other way is to analyze the synthetic observations obtained with multi-phase HI simulation and explore where the \cite{LP00} theory describing velocity crowding is applicable to describing channel maps. In what follows, we present a synergy of the approaches by analyzing thin channel maps obtained with synthetic observations of multi-phase HI, and applying the VDA both to these synthetic and actual GALFA-HI data. We explore the striation using the RHT and compare our findings with the results of earlier RHT studies. Throughout this study, we use the term "velocity caustics" to refer to HI striations that arise from turbulent velocity crowding. Notably, this term was previously used in the studies of cosmological Large Scale Structure to describe the enhancements of observed densities in the Hubble Flow as a function of redshift \citep{2007ApJ...657..262D,2009MNRAS.400.2174V}.

The paper is organized as follows. In \S~\ref{sec:data}, we describe this study's 3D numerical simulations and observational data. In \S~\ref{sec:vc}, we present numerical experiments to illustrate the velocity caustics and introduce the VDA pipeline adopted in this work. In \S~\ref{sec:VDAn} and \S~\ref{sec:vdao}, we present the results of our analysis, including numerical tests for VDA and the application of VDA to GALFA-HI data. In \S~\ref{sec:com}, we show that comparing high-intensity HI structures within thin channels and FIR emission is inappropriate for testing the velocity caustics. In \S~\ref{sec:dis}, we discuss the nature of HI striations in thin channel maps and the implications of this work for other related studies. Our results are summarized in \S~\ref{sec:con}.

\section{Numerical and Observational Data}
\label{sec:data}
\subsection{Numerical simulations}
\subsubsection{Isothermal simulations}
This study uses 3D simulations of turbulence generated through the ZEUS-MP/3D code \citep{2006ApJS..165..188H}. We briefly outline the setup for the simulations.

\begin{table}
	\centering
	\label{tab.1}
	\begin{tabular}{| c | c | c | c | c | c |}
		\hline
		Run & $M_{\rm S}$ & $M_{\rm A}$ & Resolution & Condition & Code\\\hline\hline
		Ms12MA06 & 1.2 & 0.6 & $792^3$ & isothermal& ZEUS-MP/3D\\
		Ms100MA07 & 10 & 0.7 & $792^3$ & isothermal& ZEUS-MP/3D\\\hline
		MP & 1.0 & 1.0 & $480^3$ & multi-phase & Athena++ \\\hline
	\end{tabular}
	\caption{Setups of numerical simulations. $M_{\rm S}$ and $M_{\rm A}$ are the mean values of the simulations.}
\end{table}

The ideal MHD equations, including mass density $\rho$ and velocity $\pmb{v}$, are solved with periodic boundary conditions. The equations are as follows:
\begin{equation}
	\label{eq.mhd}
	\begin{aligned}
		&\partial\rho/\partial t +\nabla\cdot(\rho\pmb{v})=0,\\
		&\partial(\rho\pmb{v})/\partial 
		t+\nabla\cdot\left[\rho\pmb{v}\pmb{v}^T+(c_s^2\rho_+\frac{B^2}{8\pi})\pmb{I}-\frac{\pmb{B}\pmb{B}^T}{4\pi}\right]=\pmb{f},\\
		&\partial\pmb{B}/\partial t-\nabla\times(\pmb{v}\times\pmb{B})=0,\\
		&\nabla \cdot\pmb{B}=0,\\
	\end{aligned}
\end{equation}
where $c_s$ is the constant sound speed due to the isothermal equation of state, and $\pmb{f}$ represents the stochastic forcing term used to drive turbulence. The magnetic field $\pmb{B}$ and density fields are initially set to be uniform, with $\pmb{B}$ aligned along the $y$-axis. 

The turbulence simulations can be characterized by the sonic Mach number ($M_{\rm S}=v_{\rm inj}/c_{s}$) and the Alfv\'{e}nic Mach number ($M_{\rm A}=v_{\rm inj}/v_{A}$) for MHD simulations, where $v_{A}=B/\sqrt{4\pi\rho}$ is the Alfv\'{e}n speed. For turbulence driving, we adopt the common driving method \citep{2003MNRAS.345..325C,2013MNRAS.436.1245F,YL17a,2020MNRAS.492..668B}. The energy injection is centered on wavenumber $k=2\pi/l=1-2$ (in the unit of $2\pi/L_{\rm box}$, where $L_{\rm box}$ is the length of simulation box) in Fourier space, where $l$ is the length scale in real space. We run the simulation for six eddy turnover times to ensure that the turbulence has reached statistical saturation. The values of $M_{\rm S}$ and $M_{\rm A}$ are listed in Tab.~1. 

\subsubsection{Multi-phase simulations}
To emulate the multi-phase ISM environment, we incorporated an additional ordinary differential equation (ODE) solver to consider radiative cooling and heating effects in addition to the original Athena++ MHD Solver. Our initial state was a 3D turbulence box with periodic boundaries and a length of 200 pc, representing bulk neutral hydrogen in the ISM. We utilized a realistic synthetic cooling and heating function proposed by \cite{2002ApJ...564L..97K} and solved the equation with an adaptive implicit solver to ensure convergence. The simulation was initially of constant density and driven by spectral velocity perturbation in Fourier space. The multi-phase medium began to form approximately 20 Myr into the simulation. To ensure a realistic representation of a multi-phase medium and accurate capture of turbulence effects, we selected a snapshot around 100 Myr. The parameters for this snapshot are listed in Tab.~ 1. For additional simulation details and statistics, see \cite{VDA} and \cite{Ho21}.

\subsubsection{Synthetic spectroscopic observations}
The observed intensity distribution of a given spectral line in PPV space is determined by both the density of emitters and their velocity distribution along the LOS. When coherent velocity shear, e.g. due to Galactic rotation (see Section~\ref{sec:Gal_rotation}) can be neglected, the LOS velocity component $v$ is the sum of the turbulent velocity $v_{\rm tur}(x,y,z)$, and the residual component due to thermal motions. This residual thermal velocity $v-v_{\rm tur}(x,y,z)$, has a Maxwellian distribution $\phi(v,x,y,z)$, so that the gas distribution in PPV cubes, $\rho_s(x,y,v)$, and in real-space. This induces intensity fluctuation in PPV that for the case of emissivity proportional to density\footnote{The quadratic dependence of emissivity is considered in \cite{2016MNRAS.461.1227K}.} provide PPV emission density $\rho_s(x,y,z)$ as \citep{LP04}:
\begin{align}
	\label{eq.max}
	\rho_s(x,y,v)&=\kappa \int \rho(x,y,z) \phi(v,x,y,z) dz,\\
	\phi(v,x,y,z) & \equiv \frac{1}{\sqrt{2\pi c_s^2}}\exp[-\frac{[v-v_{\rm tur}(x,y,z)]^2}{2c_s^2}],
\end{align}
where $\kappa$ is a constant that relates the number of emitters to the observed intensities, which is not of interest to our discussion of the mechanism of caustic formation. $c_s=\sqrt{\gamma k_{\rm B}T/m}$, with $m$ being the mass of atoms or molecules, $\gamma$ is the adiabatic index, $k_{\rm B}$ being the Boltzmann constant, and $T$ the temperature, which can vary from point to point if the emitter is not isothermal. To include the thermal line broadening effect in the synthetic PPV cubes, $\rho_s(x,y,v)$ is convoluted by a thermal Gaussian kernel of $\sigma_T=c_s$.

By integrating $\rho_s(x,y,v)$ over a given velocity range, which is called the channel width $\Delta v$, we obtain a velocity channel:
\begin{equation}
	\label{eq.p}
	p(x,y,v)=\int_{v-\Delta v/2}^{v+\Delta v/2}\rho_s(x,y,v^\prime)dv^\prime.
\end{equation} 

If we split the 3D density into the mean density and zero mean fluctuations
$\rho(x,y,z) = \bar\rho + \bar\rho \delta(x,y,z)$ we arrive at the representation of the channel intensity as the sum of two terms $p(x,y,v)=p_{vc}(x,y,v)+p_{dc}(x,y,v)$:
\begin{align}
	\label{eq:rhov}
	p_{vc}& \equiv \int_{v-\Delta v/2}^{v+\Delta v/2} \!\! dv^\prime \int \bar\rho \phi(v^\prime,x,y,z) dz, \\
	\label{eq:rhod}
	p_{dc}& \equiv \int_{v-\Delta v/2}^{v+\Delta v/2} \!\! dv^\prime \int \bar\rho\delta(x,y,z)\phi(v^\prime,x,y,z) dz.
\end{align}
The first term contains the mean intensity in the channel and carries fluctuations that are produced exclusively by velocity mapping. To reflect this we can call this term pure ''velocity'' term as far as structures in intensity maps are concerned. The second, "density", term reflects inhomogeneities in the real 3D density, however, 3D density structures are still modified by velocity fluctuations when mapped to channel intensities. 

In the case of $\Delta v = \infty$, we obtain the fully integrated intensity map $I$. The full line integrated $p_{vc}$ is reduced to the mean column density while $p_{dc}$ gives the column density variations on the POS. To generate a synthetic spectroscopic cube, we use the density field $\rho(x,y,z)$, velocity field $v_{\rm tur}(x,y,z)$, and temperature information from simulations. 
Additionally, to produce $p_{vc}$, we use a constant density field $\rho(x,y,z)$ in PPP space so that pre-existing density structures are fully excluded.

\subsection{Observational data: GALFA-HI and Planck}
In this work, we utilize the GALFA-HI survey data from Data Release 2, as described in \cite{2018ApJS..234....2P}. The HI data has a beam resolution of 4$'\times$4$'$, which has been gridded into 1$'\times$1$'$ per pixel, with a spectral resolution of $\approx0.2$ km/s and a brightness temperature noise of approximately 40~mK per 1 km s$^{-1}$ integrated channel. Our analysis covers the full velocity range.

We use also the Planck 353 GHz data from the Planck 3rd Public Data Release (DR3) 2018 of High Frequency Instrument \citep{2020A&A...641A...3P}. Observations from Planck designate the polarization angle $\phi$ with Stokes parameter maps $I$, $Q$, and $U$. Angle $\phi$ is defined  as:
\begin{equation}
	\phi=\frac{1}{2}\tan^{-1}(-U,Q),
\end{equation}
where the -U notation converts the angle from HEALPix convention to IAU convention. The Stokes parameter maps were smoothed from nominal angular resolution 5$'$ to 16$'$ with a Gaussian kernel to achieve a higher sign-to-noise ratio and match the resolution of the velocity channel gradients (see Appendix~\ref{app: vgt}). We infer the magnetic field angle from the equation: $\phi_B$ = $\phi$ + $\pi$/2.
\begin{figure*}
	\includegraphics[width=1.0\linewidth]{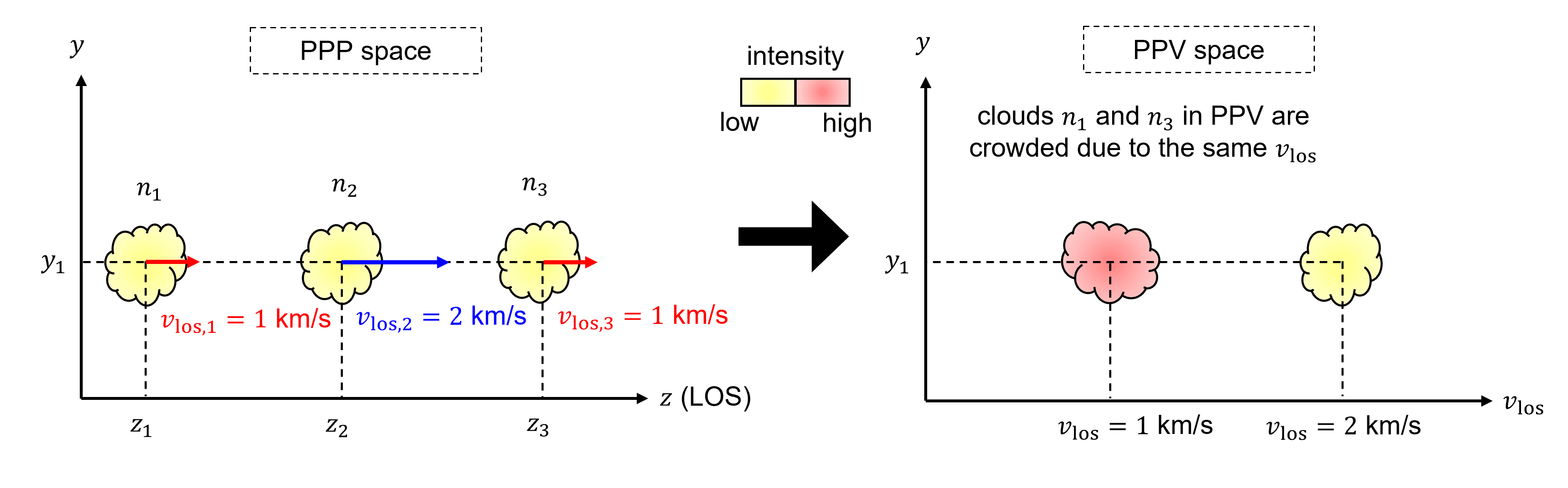}
	\caption{Illustration of velocity crowding. \textbf{Left:} three constant density ($n_1$, $n_2$, $n_3$) clouds in PPP space locate at different spatial positions along the LOS ($z$). $n_1$ and $n_3$ have identical LOS velocity $v_{\rm los}$, while $n_2$'s LOS velocity is larger. \textbf{Right:} the three clouds are mapped into PPV space. Due to $n_1$ and $n_3$'s identical LOS velocities, they are crowded into only one cloud with higher density. The cloud's morphology gets different from those in PPP space and intensity increases.}
	\label{fig:crowing}
\end{figure*}

\section{Origin of the striation: density versus velocity effects}
\label{sec:vc}
\subsection{Toy model of velocity caustics}
Fig.~\ref{fig:crowing} illustrates the velocity crowding in PPV space due to spectroscopic mapping. Three constant-density clouds ($n_1$, $n_2$, $n_3$) are located at different spatial positions along the LOS in PPP space. $n_1$ and $n_3$ possess identical LOS velocities. As a result, these clouds are mapped into the same velocity coordinate in PPV space, leading to their merging into a single crowded cloud. This new cloud is not physically present in the PPP space but is rather created by the velocity crowding effect. $n_2$ with a larger LOS velocity is mapped into a different velocity coordinate.

In realistic scenarios, the mapping process is more complex due to the turbulent and magnetized nature of the ISM \citep{1995ApJ...443..209A, 2010ApJ...710..853C, Crutcher12, BG15}. The turbulence leads to different parts of a cloud having distinct LOS velocities. Consequently, when mapped into PPV space, the cloud is substantially distorted, resulting in multiple HI clouds or sparsely distributed HI gas along the same LOS becoming crowded in PPV space. This crowding redistributes the HI gas in the PPV space and produces new HI intensity structures.

\begin{figure*}
	\includegraphics[width=1.0\linewidth]{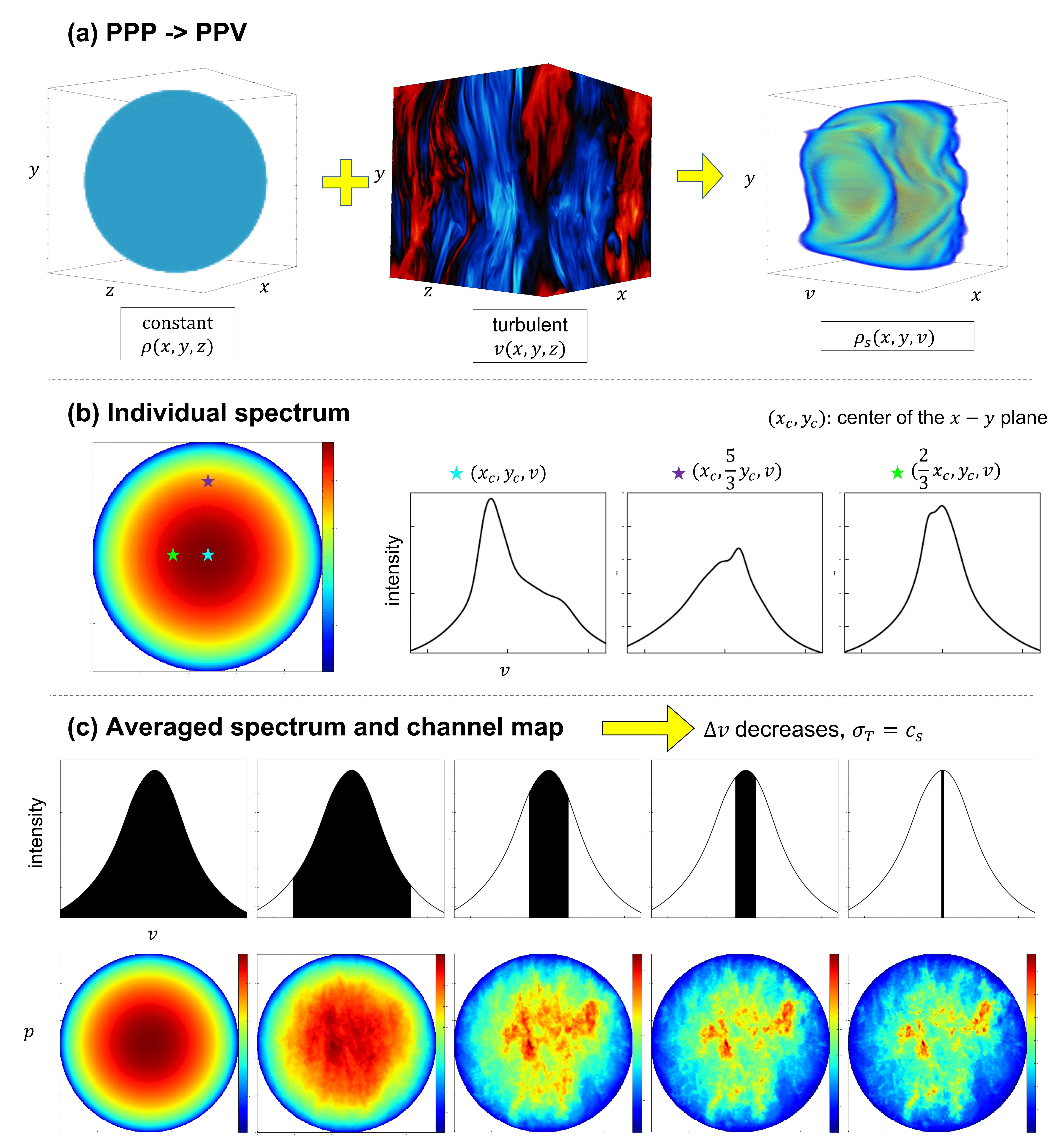}
	\caption{\textbf{Panel (a):} numerical 3D visualization of velocity caustics. The PPV cube $\rho_s(x,y,v)$ is generated with an input density field $\rho(x,y,z)$ and velocity field's LOS component $v(x,y,z)$ in PPP space. The density field $\rho(x,y,z)$ is a constant sphere, while the LOS velocity component $v$ is adopted from the multiphase simulation. The mean magnetic field is oriented along the $y-$direction. \textbf{Panel (b):} spectral (three sub-plots on the right) of the synthetic PPV cube at three different positions on the $x-y$ plane. The three positions are labeled on the (left) integrated intensity map with star symbols. \textbf{Panel (c):} the velocity spectra (top) and channel maps $p$ (bottom) of the PPV cube presented in panel (a). The velocity range used for integration (i.e., channel width $\Delta v$) is indicated by the shaded region in the top spectra (averaged over the full cube). Thermal broadening with a Gaussian kernel $\sigma_T=c_s$ is included. }
	\label{fig:vc}
\end{figure*}

Fig.~\ref{fig:vc} presents a synthetic example of PPV cubes $\rho_s(x,y,v)$ generated from a density field $\rho(x,y,z)$ and the LOS component of the velocity field $v(x,y,z)$, accounting for thermal broadening effects. The density field $\rho(x,y,z)$ is a constant sphere, while the LOS velocity component $v$ is derived from multiphase HI simulations. We can see the sphere in PPV space are considerably distorted. The constant $\rho(x,y,z)$ eliminates any pre-existing density structures that are correlated with the magnetic field so that the projected $\rho(x,y,z)$ map appears only as a circular structure (see Fig.~\ref{fig:vc}, panel b). Moreover, we take spectral lines along three different LOS. The lines' amplitude is not constant, but varies as a function of velocity, resembling the line profiles obtained in observations. Then, in Fig.~\ref{fig:vc}, panel (c), we take channel maps by using the spectral line averaged over the cube and vary the channel width (i.e., the velocity range used for integration). We can see a fully integrated channel recovers the column density map, erasing velocity information. The significance of crowding's significance is related to the channel width $\Delta v$. The velocity crowding effect gradually becomes more pronounced as the channel width decreases. In a thin channel map, the intensity structures become filamentary. These intensity structures are solely created by the velocity crowding effect.

\subsection{Effect of the Galactic rotation}
\label{sec:Gal_rotation}
The LOS velocity in the ISM contains a contribution from Galactic rotation. It is commonly assumed that velocity information in spectroscopic PPV space can be used to determine the Galactic spatial location of observed objects using the Galactic rotation curve. However, the extent to which this is possible and over what distance scale it is possible is determined by comparing the LOS projected shear of Galactic rotational velocity with turbulent velocity over the same distance. 

\cite{LP00} noted that in the vicinity of the Galactic plane, the coherent rotational shear is approximately 15~km/s per kpc of separation (as given by Oort's constants) and is even smaller in high-latitude regions. The turbulent velocity, on the other hand, is typically around 10~km/s at 100 pc separation \citep{2022ApJ...934....7H} and scales as the cubic root of separation $l$, i.e., $v_{\rm tur}=v_{\rm inj}(l/L_{\rm inj})^{1/3}$. Therefore, at the injection scale $L_{\rm inj}\sim100$~pc to which turbulent cascade may extend, rotational velocity differences are only 1.5~km/s, which is one-sixth of turbulence injection velocity. Recovering LOS positions of HI parcels from the rotational curve within such distances in the presence of turbulent motions is very challenging. In this paper, we only briefly discuss the effect of Galactic rotation. One can refer to \cite{LP00} for details of the effect of rotational motion on velocity mapping.

\subsection{Effect of thermal broadening}
Understanding the formation of velocity caustics in the presence of shear and thermal broadening can be achieved by considering the three settings below.
\begin{itemize}
	\item \textbf{Setting 1.} Emitting atoms with zero thermal and turbulent velocities produce emission in one channel centered at $v=0$. In this case, the channel thickness does not affect the observed intensities, and all the fluctuations arise from density fluctuations. If density filaments are present, their projected images are seen.
	\item \textbf{Setting 2.} Emitting atoms with thermal velocities and  temperatures $T$ produce emission in the range $\Delta v\sim \sqrt{k_B T/m}$, and all fluctuations arise from densities. When channels are thinner than $\sqrt{k_B T/m}$, the intensity of channel maps changes because only part of the total emission intensity of atoms (which is proportional to column density) is included in the range $\Delta v < \sqrt{k_B T/m}$. The change in channel width induces changes in intensity, so the intensity of fluctuations measured in a channel is correlated to the channel's thickness. However, with only thermal velocity considered, the morphological pattern of the intensity fluctuations in the velocity channel maps resembles that of the total integrated intensity, as well as column density (see Eq.~\ref{eq.max}). If a real density filament exists in this setting (i.e., with only thermal velocity), its projection on the POS can also be observed in the channel map.
	\item \textbf{Setting 3.} Emitting atoms have both thermal and turbulent velocities, leading to three sub-cases:
	\begin{itemize}
		\item (1). The regular velocity with shear only. In this case, the velocity mapping provides a proxy for the third dimension, potentially providing insight into the emitter's distribution along the LOS. The intensity distribution observed in velocity channels can change\footnote{As we discussed in \S~\ref{sec:Gal_rotation}, the turbulent velocity shear increases with the decrease of the scale, the effects of shear arising from Galactic rotation has little effect on velocity caustics at small scales.}.
		\item (2). Turbulent and thermal velocities only, no density fluctuations (i.e., emitting atoms uniformly distributed in PPP space). The turbulent velocities create intensity fluctuations, i.e., velocity caustics, in PPV space, as illustrated in Fig.~\ref{fig:crowing}. Due to thermal broadening, the intensity fluctuations raised from turbulent velocities in a channel might be erased or dominated by those raised by thermal velocities. Their relative significance depends on the temperature and channel width. For a given channel width, a higher temperature means a stronger thermal effect. However, when the channel width decreases, the significance of velocity caustics increases, so caustics are seen in thin channel maps (see Fig.~\ref{fig:sub09}).
		\item (3). Both velocity and density contributions affect the intensity distribution in channel maps, but their relative importance, $p_v$ and $p_d$, varies with the thickness of the channel map. When a thick channel is integrated over a large velocity range, such as the entire velocity range, only a column density map with no velocity information is obtained (see Fig.~\ref{fig:vc}). However, as the channel width decreases, the contributions from densities decrease while the velocity contributions increase. When the channel width is less than the turbulence velocity dispersion, $\sigma_v$, the velocity contribution dominates over the density contribution, provided that the mean density is greater than the density fluctuation (see \citealt{LP00} for a more detailed discussion). Note that in any case, the $p_d$ term in the thin channel regime is not a pure spatial density perturbation but is also modified by velocity mapping.
	\end{itemize}
\end{itemize}

Settings 1 and 2 are straightforward, while Setting 3 (1) is widely discussed in the literature as a means of mapping the 3D distribution of emitters using the Galactic rotation curve. However, a caveat of this setting is the absence of turbulent velocities. In the presence of turbulence, a narrow velocity channel may not accurately represent the actual density distribution along the LOS. Nevertheless, the turbulence effect can be mitigated by increasing the thickness of the channel maps. Furthermore, as indicated in Setting 3 (2) and (3), we know that (i) the thermal effect can decrease the velocity contribution in a channel, and (ii) the relative significance of velocity caustics depend on the thickness of the channel maps. In Figs.~\ref{fig:sup04} and \ref{fig:sub09}, we showed the thermal effect at a normal level and density contribution are sub-dominated in thin channels using isothermal MHD simulations. Nevertheless, in the following, we introduce how to separate the thermal effect and $p_d$ from a channel.

\subsection{Separating velocity contribution and removing thermal broadening}
\label{subsec:vda}
To understand the origin of the elongated HI structure within thin channels, it is crucial to identify a suitable and dependable way to separate the velocity and density contribution, as well as remove thermal broadening, in PPV channels. Then one can directly test the velocity caustics effect. 
Theoretical separation is given by Eqs.~(\ref{eq:rhov}, \ref{eq:rhod}) that generates a pure velocity caustics contribution, $p_{vc}$ can be accomplished in simulations by using a constant density field $\rho(x,y,z)$ in PPP space and simulated velocity $v(x,y,z)$. It is, however, difficult to apply directly to observational channel data, since the mean spatial density is unknown. To overcome this difficulty, \citet{VDA} introduced a novel technique named Velocity Decomposition Algorithm (VDA). \footnote{The effectiveness of the VDA has been thoroughly tested using multi-phase HI simulations and GALFA-HI data \citep{VDA,Yuen23}. While the validity of the VDA was strongly questioned by \cite{2022arXiv220201610K}, this paper was based on the error in the understanding of the VDA. This was pointed out in \cite{2022arXiv220207871Y}, and the aforementioned authors retracted their paper from $A$\&$A$. } In this paper, we discuss the VDA only briefly, with additional details to be found in \citet{VDA}. 

From the intensity distribution within a channel, denoted as $p(x,y,v)$, VDA extracts the velocity $p_v(x,y,v)$ and density $p_d(x,y,v)$ contributions in the channel according to the following prescription:
\begin{equation}
	\begin{aligned}
		p_d&=\left(\langle p\cdot I\rangle-\langle p\rangle\langle I\rangle\right)\frac{I-\langle I\rangle}{\sigma_I^2},\\
		p_v&=p-p_d = p-\left(\langle p\cdot I\rangle-\langle p\rangle\langle I\rangle\right)\frac{I-\langle I\rangle}{\sigma_I^2},
	\end{aligned}
\end{equation}
where $I \equiv \int_{-\infty}^\infty dv p(x,y,v)$, $\sigma_I^2=\langle(I-\langle I\rangle)^2\rangle$ and $\langle...\rangle$ denotes the ensemble average over the entire map~\footnote{In observation, $p$ and $I$ can be obtained from:
	\begin{equation}
		\label{eq.pI}
		\begin{aligned}
			p(x,y,v)&=\int_{v-\Delta v/2}^{v+\Delta v/2}T_{\rm mb}(x,y,v^\prime)dv^\prime,\\
			I(x,y)&=\int_{-\infty}^{+\infty}T_{\rm mb}(x,y,v^\prime)dv^\prime,
		\end{aligned}
	\end{equation} 
where $T_{\rm mb}$ is the brightness temperature.}. Defined this way, $p_d$ describes the channel intensity fluctuations proportional to column density fluctuations at the level given by the cross-correlation coefficient between channel and column intensities. The "velocity" part $p_v$ contains the contribution from the pure velocity part $p_{vc}$.
	
In addition, it is worth noting that $p_v$ and $p_d$ have the following fundamental properties:
\begin{equation}
	\begin{aligned}
		\langle p_d \rangle &= 0 ,\quad  \int_{-\infty}^\infty \! dv \; p_d(x,y,v) = I-\langle I\rangle,\\
		\langle p_v \rangle &= \langle p \rangle, \quad \int_{-\infty}^\infty \! dv \;p_v(x,y,v) = \langle I \rangle, \\
		\langle p_v p_d \rangle &= 0,
	\end{aligned}
\end{equation}
which highlights firstly the fact that $p_d$ is a perturbative term, carrying information only about density inhomogeneities, while $p_v$ contains the mean intensity background and velocity field contribution, and secondly the orthogonality of VDA decomposition in the sense that $p_d$ and $p_v$ are uncorrelated.

The values $p_{v}$ are available directly via the application of VDA to observations. On the contrary, the quantity $p_{vc}$ is only available using numerical simulations with a constant density and serves as a theoretical template for the structures generated purely by velocity mapping.  The value of the chosen constant density rescales $p_{vc}$. Thus, if the density is significantly non-uniform, the magnitude of $p_{vc}$ will depend on which volume the mean density is determined. For instance, if one studies the small scale structure of a cloud, the appropriate value of the mean density entering velocity mapping is that of a cloud, rather than the lower value averaged over wider emptier space. This rescaling of $p_{vc}$, however, does not change the shape of the structures that this part of the signal exhibits.

Respectively, it is important to acknowledge the limitations inherent in the VDA approach. The VDA model assumes the presence of a single emitting turbulent region with velocity and density fluctuations. While this assumption is accurate for a turbulent cloud, such as a localized molecular cloud, it becomes problematic when dealing with HI gas that consists of multiple components along the LOS. In such cases, the application of VDA would result in the cumulative density of all the components being represented by $p_d$. To mitigate this issue, the VDA can be applied to the emission line decomposed into Gaussian components, as described in \cite{2022MNRAS.513.3493H}. Furthermore, the accuracy of determining $\langle I \rangle$ can be affected by an inhomogeneous intensity distribution across the analyzed image. To address this, it is possible to filter out low spatial frequencies of the image or apply the VDA to sub-blocks of data. These improvements are particularly relevant when applying the VDA to HI in the Galactic disk that exhibits multiple peaks. In the present study, we include three clouds in both high and low galactic latitudes. This allows us to emphasize the importance of considering velocity caustics in our analysis.

Bearing this in mind, we proceed to analyze a multi-phase simulation of magnetized HI, which will enable us to determine whether the striation observed in channel maps arises from velocities or densities.

\section{Numerical testing: multi-phase HI simulations}
\label{sec:VDAn}

\begin{figure*}
	\includegraphics[width=1.0\linewidth]{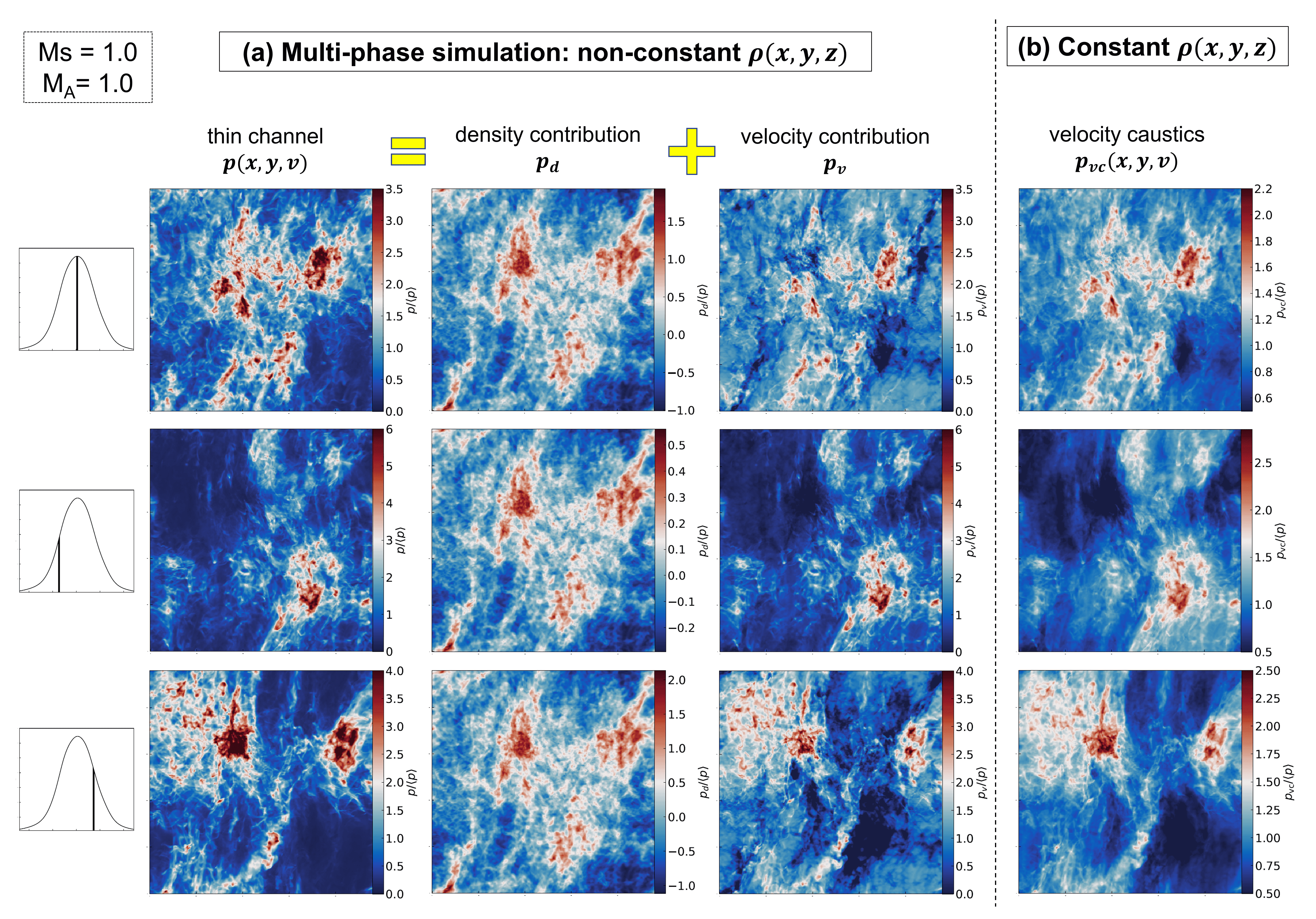}
	\caption{Schematic of the formation of images velocity channel maps and their comparison with velocity caustics. \textbf{Panel (a)}: Results of the separation of density contribution ($p_d$) and velocity contribution ($p_v$) from a thin channel ($p$) using a non-constant 3D density field $\rho$ from the MHD simulations. The velocity range used for integration is indicated by the shaded region in the top spectrum. The color bars for panels provide the relative contributions. The left graphs show the position of the channels relative to the average profile of the line \textbf{Panel (b)}: The thin channel $p_{vc}$ in this panel was generated using a constant density field equal to the full volume mean density, thereby eliminating any pre-existing density structures. The structures within the thin channel are solely created by velocity mapping. The integration velocity range is identical to that in panel (a). The mean magnetic field is oriented along the vertical $y-$direction and thermal broadening is included. The channel width $\approx0.2$~km/s is selected to match the GALFA-HI data.}
	\label{fig:mp}
\end{figure*}

\begin{figure*}
	\includegraphics[width=1.0\linewidth]{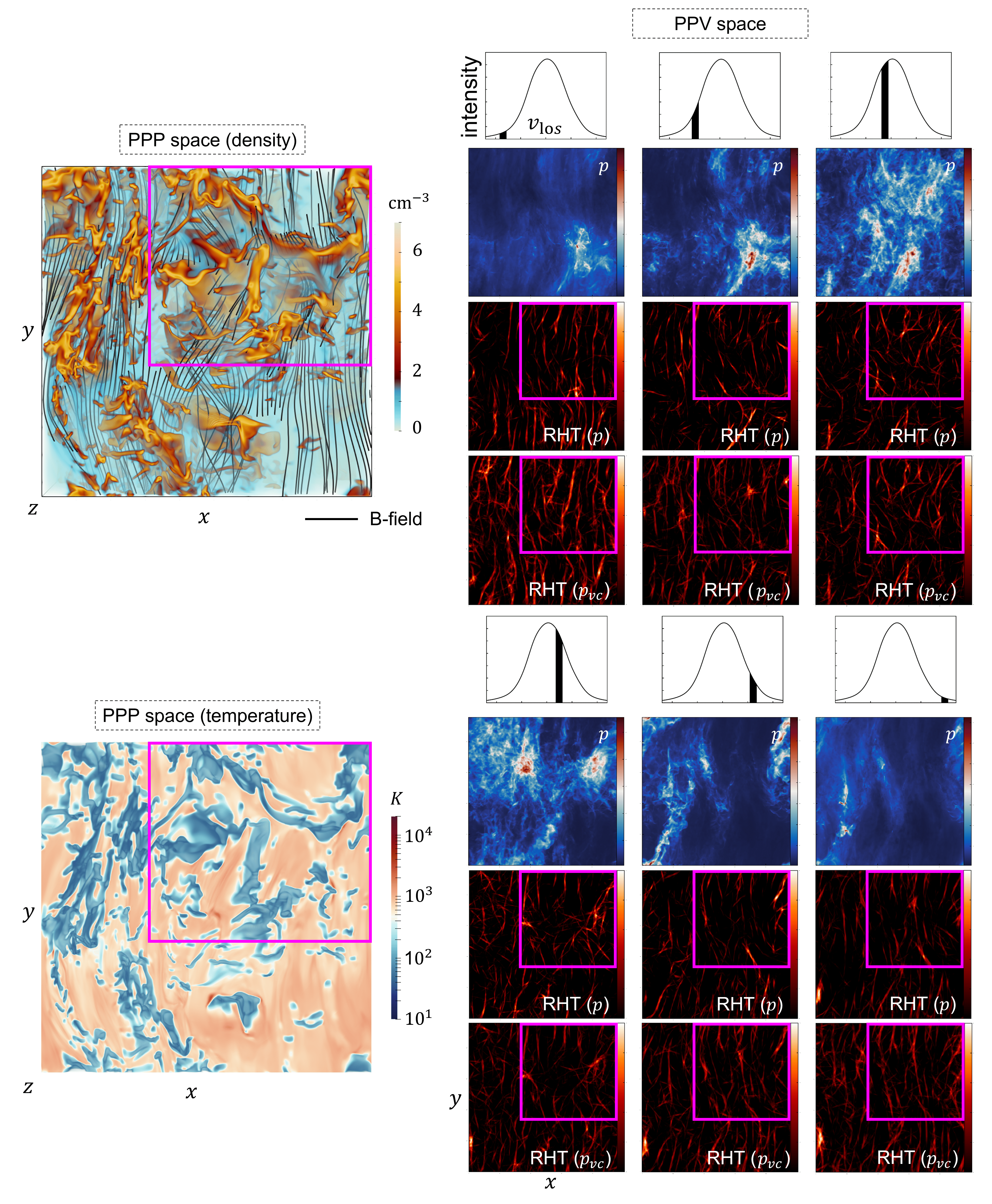}
	\caption{\textbf{Left}: 3D visualization of the multi-phase simulation's volume density field and magnetic field (top), as well as temperature (bottom), in PPP space. In the highlighted square box specifically, one observes density structures that are mostly orthogonal to the magnetic field. \textbf{Right}: synthetic thin ($\Delta v\approx1.5$~km/s) channel maps $p$ generated from the multi-phase simulation and their corresponding RHT-fibers. The RHT-maps are calculated from $p$ and their corresponding $p_{vc}$ maps, which are generated from a constant density field. Thermal broadening is included. The shaded area at the top spectra indicates the velocity range used for integration.  }
	\label{fig:mp3d}
\end{figure*}

\subsection{Multiphase HI simulations}
\label{subsec:MP}
\subsubsection{Tests for VDA}
In this section, we apply VDA to synthetic observation with the purpose of determining the relative level of "density" $p_d$ and $p_v$ contributions as well as analyzing the level at which $p_v$ can be explained by the pure velocity caustic effect as encoded in $p_{vc}$. In simulations, $p_{vc}$ is the result of using simulation velocities and applying them to constant density (as in Fig.~\ref{fig:vc}). At the same time, $p_v$ is the result of VDA separation of the effect of velocity caustics from the effect of density fluctuation.  A numerical test of VDA's validity in removing thermal broadening and separating velocity contributions using isothermal MHD simulations is given in Fig.~\ref{fig:sub09}. In the main text, we directly start with the multi-phase simulation. More on the VDA justification and application can be found in \cite{VDA}. 

In Fig.~\ref{fig:mp}, we perform VDA in $M_{\rm S}=1.0, M_{\rm A}=1.0$ multi-phase HI MHD simulation and compare obtained $p_v$ and $p_d$ with the pure velocity contribution $p_{vc}$ and the full channel intensity $p$. Different from isothermal MHD simulations, the multi-phase simulation is not scale-free and is composed of the cold neutral medium (CNM), unstable neutral medium (UNM), and warm neutral medium (WNM). The comparison is performed for three channels centering at different $v_{\rm los}$. The channel associated with the most prominent intensity is denoted as the central channel and the other is called the wing channel.

We can see that for both the central and wing channels, intensity structures in $p$ are similar to those in velocity contribution $p_{v}$ and velocity caustics $p_{vc}$. This means velocity caustics has an important contribution to channel maps. The perturbative quantity $p_d$ is
allowed to have negative values and its amplitude varies because $p_d$ contains the information of density fluctuations in a channel. Nevertheless, the difference between central and wing channels exists. For the central channel, the similarity also appears between $p$ and $p_d$, while this is not observed in the wing channel. It suggests $p_d$ has more contribution to the central channel, while $p_v$ is still dominated. In both cases, the correspondence of $p_v$ and the velocity caustics $p_{vc}$ tests the performance of VDA with synthetic observations of multi-phase HI. In short summary, Fig.~\ref{fig:mp} testifies that: 
\begin{itemize}
	\item The contributions to channel $p$ arising from density $p_d$ and velocity $p_v$ are different. 
	\item Structures observed in thin velocity channels $p$ are dominated by velocity contribution $p_{v}$. The effect of density variations $p_d$ is marginal for thin velocity channels, especially the wing channel.
	\item The correspondence of the $p_v$ and $p_{vc}$ is very good (quantification is given in \S~\ref{subsub:ncc}). Thus, $p_v$ obtained via VDA can be used as a proxy for $p_{vc}$ where the information on the mean density is unavailable and VDA provides a tool for separating density and velocity contributions to intensity striations observed in the PPV channel, irrespective of the thermal effect's significance (see Fig.~\ref{fig:sub09}).
\end{itemize}
We discuss the application of VDA to actual HI observational data in \S~\ref{sec:vdao}.

\subsubsection{Nature of striations in thin channels}
The 3D visualization of the multi-phase HI simulation's density cube in PPV space is given in Fig.~\ref{fig:mp3d} (left side). It shows many high-density filaments can be aligned either parallel or perpendicular to the magnetic field that is shown by black lines. The comparison between the upper and lower left panels shows that the dense structures correspond to cold gas. The highlighted magenta squares show that the magnetic field is mostly vertical, and the cold density structures/filaments are preferentially horizontal. This confirms that PPP cold-density filaments are not necessarily parallel to magnetic fields which corresponds well to theoretical expectations \citep{2019ApJ...878..157X}. Similarly, these (perpendicular) cold density filaments are also observed in \cite{2023arXiv230504965G} using multi-phase simulations (see their Fig.~10).

To objectively analyze the velocity caustics' impact on shaping the intensity distribution in PPV space, we plot six thin channel maps ($\Delta v\approx1.5$~km/s) generated from the multi-phase simulation and located at different velocity coordinates, covering almost the entire velocity range. In all these channel maps, we observe the intensity structures preferentially follow the magnetic field. This is also true for the regions where the cold-density structures tend to be perpendicular to the magnetic field. This testifies that the density contributions are subdominant to what is seen in the channel maps.

To better visualize the striated intensity structures in thin channel maps, we follow \cite{Clark15} and employ the Rolling Hough Transform (RHT; \citealt{2014ApJ...789...82C,Clark15}), which is an algorithm for extracting linear structures in an image. The RHT results depend on the somewhat arbitrary input parameters \citep{2014ApJ...789...82C} of smoothing kernel diameters (DK), a window diameter (DW), and an intensity threshold (Z). Different parameters may affect the degree of alignment of "fibers" with the magnetic field (see Figs.~5, 6, 7 in \citealt{2014ApJ...789...82C}). We repeated the RHT analyses with different parameters and selected the "visually correct" result with parameters of DK = 11, DW = 55, and Z = 0.7. For this choice of parameters, the RHT provides a better tracing of the magnetic fields. 

In Fig.~\ref{fig:mp3d} (right side), we present the RHT-identified structures (i.e., RHT-fibers) for the six channel maps. Looking at the magenta squares it is easy to see that the cold-density filaments perpendicular to the magnetic fields seen in PPP space are not seen in thin channel maps. The RHT-identified filaments are aligned parallel to the projected magnetic field. Using the synthetic $p_{vc}$ maps obtained with constant density, we observe that RHT-fibers obtained for $p_{vc}$ are nearly identical to the fibers thin channel maps $p$. This proves that RHT-fibers in thin channel maps arise from velocity caustics. The alignment of RHT-fibers with magnetic field in our synthetic observations is similar to that was observed earlier in GALFA-HI data (see \citealt{2014ApJ...789...82C,Clark15}). This suggests that striations of intensity observed in GALFA thin channel maps are dominated by velocity caustics rather than cold-density PPP filaments. Further in the paper (see \S 5.2), we strengthen this conclusion by applying VDA to GALFA data.

Note that our analysis does not mean that cold HI filament cannot be aligned with the magnetic field, but shows that the contribution of these filaments is subdominant for RHT-filaments detected in thin channel maps. Indeed, if we examine PPP filaments beyond the magenta square, we observe cold filaments parallel aligned with magnetic fields, as seen in Fig.~\ref{fig:mp3d}. Such alignment is expected in MHD turbulence theory \citep{2019ApJ...878..157X,2021MNRAS.504.4354B} e.g., if turbulent velocities passively advect the density fluctuations. This is, for instance, the case for density inhomogeneities arising from entropy fluctuations when turbulent gas is not isothermal. In this case, the statistics of density fluctuations mimic the statistics of the velocity \citep{monin2013statistical}, and therefore the parallel alignment of density filaments is expected in PPP space. Thus, the velocity field is responsible for the parallel (to magnetic field) alignment of both RHT-fibers in thin channel maps and the occasional alignment of cold-density filaments.\footnote{When the back reaction from the gas is important, the parallel to magnetic field alignment is not enforced, in agreement with our simulations.}

In other words, the correspondence of the alignment of filaments observed in high Galactic latitude FIR maps and the RHT-detected fibers in channel maps (see Fig.~1 in \citealt{2019ApJ...886L..13P}, Fig.~8 in \citealt{Clark19}, and Fig.~9 in \citealt{2020arXiv200301454K}) cannot be used as an argument that the RHT fibers are images of PPP cold density filaments. In \S~\ref{sec:dis}, we present another example: FIR filaments are perpendicular to magnetic fields.  

\begin{figure*}
	\includegraphics[width=0.99\linewidth]{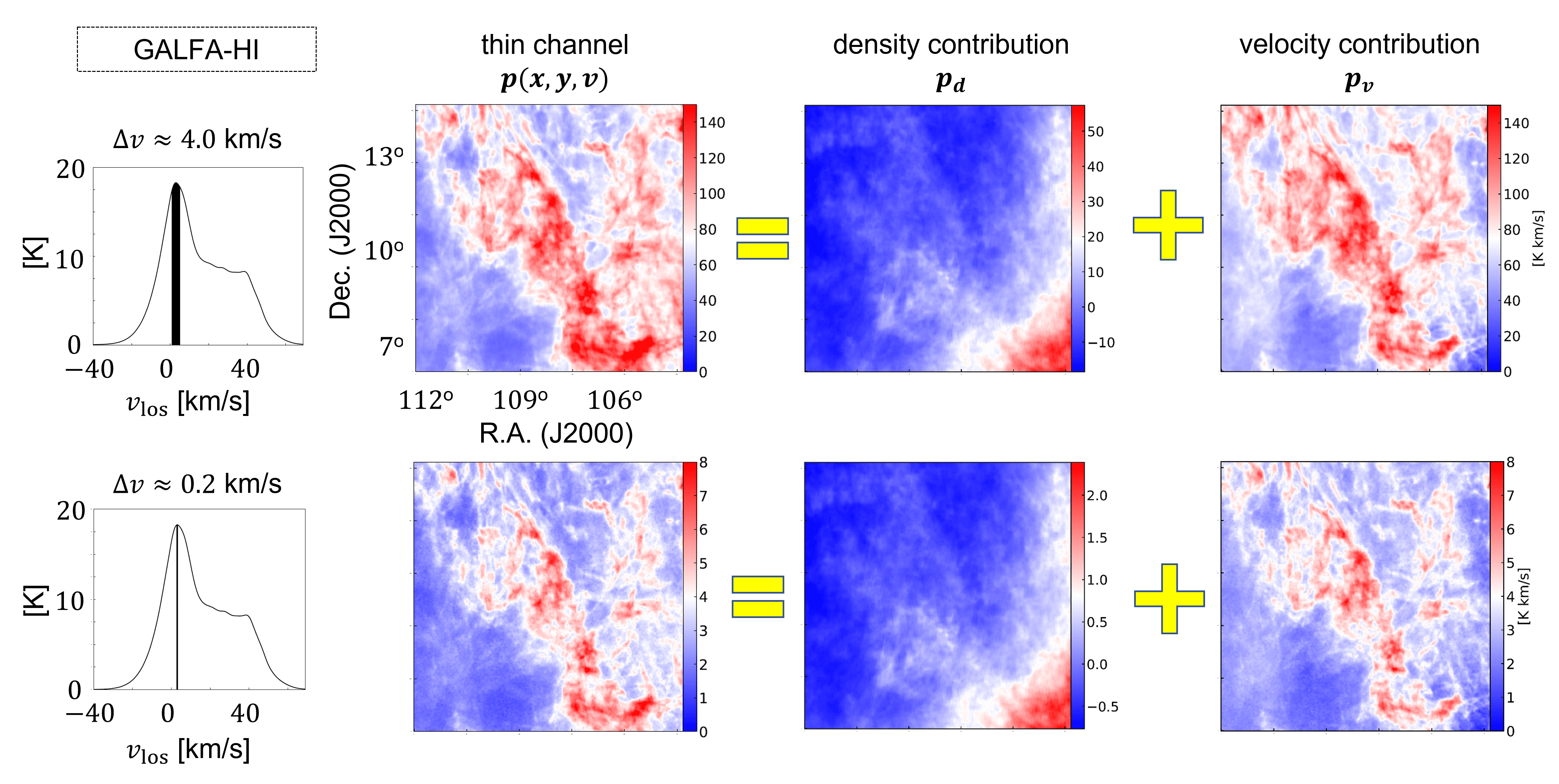}
	\caption{Same to Fig.~\ref{fig:mp}, but for a HI cloud ($v_{\rm los}\approx3$~km/s) selected from the GALFA-HI survey. The shaded area at the top indicates the velocity range (i.e., channel width $\Delta v$) used for integration. Two different channel widths have been adopted: $\Delta v\approx4.0$ km/s (top) and $\Delta v\approx0.2$ km/s (bottom).}
	\label{fig:galfa_vda}
\end{figure*}

In summary, our analysis of synthetic observations obtained with multi-phase HI reveals:
\begin{itemize}
	\item Some PPP cold-density filaments in multi-phase HI are  perpendicular, while others are parallel to the ambient magnetic field.  
	\item These cold-density filaments marginally affect the striations observed in thin velocity channel maps.
	\item RHT applied to thin channel maps delivers striations that arise from velocity caustics, and these RHT-fibers are parallel to the magnetic field.
\end{itemize}

\section{VDA's application to HI observational data}
\label{sec:vdao}
\subsection{Dominance of velocity caustics in thin channels}
\label{subsub:ncc}
In \S~\ref{sec:VDAn}, we presented a simulation-based visual comparison between the VDA-separated velocity contribution $p_v$ in a thin channel and the pure velocity caustics channel. In this section, we provide an observational example using GALFA-HI data and quantify the significance of velocity caustics.

Fig.~\ref{fig:galfa_vda} displays two channel maps of the HI cloud, located at $v_{\rm los}\approx 3$~km/s but with different channel widths ($\Delta v \approx 4.0$~km/s and $\Delta v \approx 0.2$~km/s). Noticeable differences can be observed between $p_d$ and $p_v$. In addition, $p_d$ associated with low intensity has only a marginal contribution to the high-intensity filamentary striations observed in $p$. On other the hand, structures seen in $p$ are highly similar to those in $p_v$, in terms of topology and intensity amplitude. This suggests that both channels ($\Delta v \approx 4.0$~km/s and $\Delta v \approx 0.2$~km/s) are dominated by prominent velocity caustics. Two more examples at different Galactic latitudes are given in Fig.~\ref{fig:vda2}.

\subsubsection{NCC analysis}
We further use the normalized covariance coefficient (NCC) to quantify the correlation between $p_v$ and other maps. NCC of two maps $A$ and $B$ is defined as \citep{Yuen19}:
\begin{equation}
	{\rm NCC}=\frac{\langle(\langle A - \langle A\rangle)(\langle B - \langle B\rangle)\rangle}{\sigma_A \sigma_B},
\end{equation}
where $\sigma_A$ and $\sigma_B$ represent the standard deviation of maps $A$ and $B$, respectively. NCC ranges from -1 to 1, with NCC = 1 indicating that the two maps $A$ and $B$ are statistically perfectly correlated. NCC = 0 and NCC = -1 correspond to the two maps being statistically uncorrelated and anticorrelated, respectively. NCC focuses on the comparison of the structures' morphology in two maps, rather than on their relative magnitudes, to which the NCC coefficient is insensitive.

\begin{figure*}
	\includegraphics[width=0.9\linewidth]{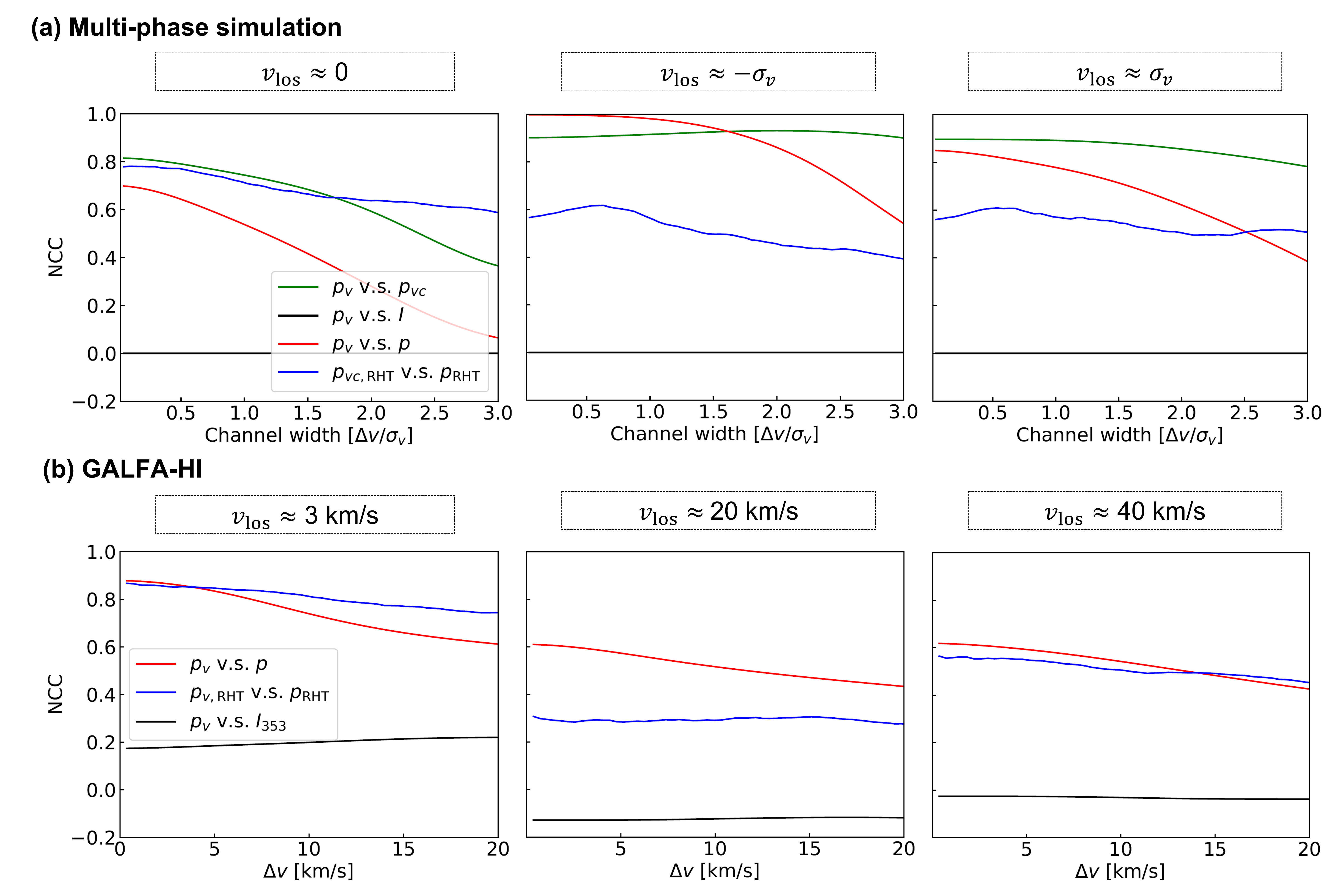}
	\caption{\textbf{Panel (a)}: This panel displays the dependence of NCC on channel width $\Delta v$, normalized by the velocity dispersion $\sigma_v$. NCC values range from -1 to 1, where 1 represents a perfect correlation, 0 indicates no correlation, and -1 represents anti-correlation. NCC is calculated between the velocity contribution $p_v$, thin channel maps $p$, column density map $I$, and pure velocity caustics map $p_{vc}$. The column density map $I$ is integrated along the full LOS velocity range to erase velocity information. $p_{\rm RHT}$ and $p_{vc,{\rm RHT}}$ are the RHT-processed $p$ and $p_{vc}$ maps, respectively. \textbf{Panel (b)}: This panel presents the NCC values between the $p_v$, $p$, and Planck 353 GHz FIR intensity map $I_{353}$ for the GALFA-HI cloud (see Fig.~\ref{fig:galfa_vda}). The Planck map only contains density information in the cold phase. $v_{\rm los}$ refers to the central LOS velocity of the channels. $p_{v,{\rm RHT}}$ is the RHT-processed $p_{v}$ map.}
	\label{fig:ncc}
\end{figure*}

Fig.~\ref{fig:ncc} displays the correlation between normalized NCC and channel width for the multi-phase simulation, as well as the GALFA-HI cloud. 
We start with the case of the simulated central channel associated with the most prominent intensity. The numerical comparison reveals a strong correlation between $p_v$ and $p_{vc}$, with NCC$\sim0.8$ when the channel is \textit{thin}, $\Delta v/\sigma_v<1$ \citep{LP00}.  Given that $p_{vc}$ is generated from a constant density field and contains solely velocity information, this high NCC value implies that $p_v$ reveals velocity structures. Concurrently, a strong correlation ($ \mathrm{NCC} \sim 0.7 $) is detected between $p_v$ and $p$, contrasting with the absence of any correlation with the column density $I$. The correlation between $p_v$ and $p$ diminishes as the channel width increases, an expected outcome since a thicker channel integrates more density contributions.
	
To better understand the origin of RHT-fibers, we employ RHT to process $p_{vc}$ and $p$ maps using the same parameters as those in Fig.~\ref{fig:mp3d}, resulting in $p_{vc,{\rm RHT}}$ and $p_{\rm RHT}$ respectively. As shown in Fig.~\ref{fig:ncc}, $p_{vc,{\rm RHT}}$ and $p_{\rm RHT}$ always display a strong correlation with $\mathrm{NCC} >0.6$ regardless of the channel width. Considering $p_{vc,{\rm RHT}}$ contains only velocity information, this confirms that the shapes of the observed RHT-fibers in $p_{\rm RHT}$ to a large extent follow velocity caustics.

In addition to the central channel, we repeat the NCC analysis for two more channels centering at $v_{\rm los}\approx\pm\sigma_v$, called wing channels. We found that NCC values of $p_{v}$ v.s. $p_{vc}$ and $p_v$ v.s. $p$ in the wing channels are higher than in the central channel, even achieving $\mathrm{NCC} >0.9$ when $\Delta v/\sigma_v<1$. This would suggest that the velocity contribution is more significant in wing channels. This effect, however, is in part due to large scale motions which separate high velocity regions into different channels. When we filter out the large scale modes in RHT analysis, we find  the correlation between $p_{vc,{\rm RHT}}$ and $p_{\rm RHT}$ in wing channels, while still rather strong, $ \mathrm{NCC} \sim 0.6$, to be somewhat lower than in the central channel.  
	
Next, we calculated the NCC for the GALFA-HI cloud at three different $v_{\rm los}\approx 3$, $20$, and $40$ km/s. Here, with $p_{vc}$ not available, we limit ourselves with comparing $p_v$ to the raw channel $p$ and the Planck 353 GHz FIR map $I_{353}$. As Fig.~\ref{fig:ncc} shows, $p_v$ is highly correlated with $p$ when the channel is narrow. In particular, NCC exceeds $0.6$ when $\Delta v \approx 0.2\; \mathrm{km/s}$. In the central channel, NCC values are as high or higher as in the simulations, however in the wings they are lower. The processed $p_{v,{\rm RHT}}$ maps also show a high correlation with the $p_{\rm RHT}$ in the central channel, at the same level of $\mathrm{NCC} \sim 0.8$ as $p_{vc,\mathrm{RHT}}$ showed in the simulations. This provides observational evidence that RHT-fibers here originate from velocity caustics. However in the wings $p_{v,{\rm RHT}}$ behaviour is less consistent than that of a simulated $p_{vc,\mathrm{RHT}}$. We expect that the existence of multiple HI clouds together with shear velocity makes it challenging to interpret wing channels, which in simulations are defined for a single HI cloud. In addition, we have replicated our analysis for two more clouds, including a high-latitude one (see Appendix~\ref{app: vda}). These clouds exhibit still high NCC values for $p_v$ versus $p$, hinting at the prevalence of velocity caustics.

Additionally, the NCC values between $p_v$ and $I_{353}$ range from -0.2 to 0.2, suggesting an insignificant correlation. The value is not strictly zero since $I_{353}$ is not the column density exactly. A similar comparison of $p$ and the Planck FIR map was conducted in \cite{Clark19}. However, they used an unnormalized parameter to quantify the correlation. This correlation was proven to be insignificant in later studies (see Appendix~\ref{app: I857} and \citealt{Yuen19}).

\subsection{RHT fibers in thin channels originate from velocity caustics}
\label{sec:RHTfil}
In \S~\ref{subsec:MP}, we used the synthetic channel maps obtained with simulations of magnetized multi-phase HI to demonstrate that the RHT-fibers in the thin velocity channel maps arise from velocity caustics. Below we gauge the relative effect of velocity and density contributions to RHT-fibers in GALFA channel maps. This study is important as the striations of HI structures within thin channels processed with RHT were proposed to trace the magnetic fields \citep{2014ApJ...789...82C,Clark15}. That study, however, explained the RHT-fibers in channel maps as images of PPP cold density filaments \citep{Clark19}. Thus it is important to put this explanation to the test.

\begin{figure*}
	\includegraphics[width=1.0\linewidth]{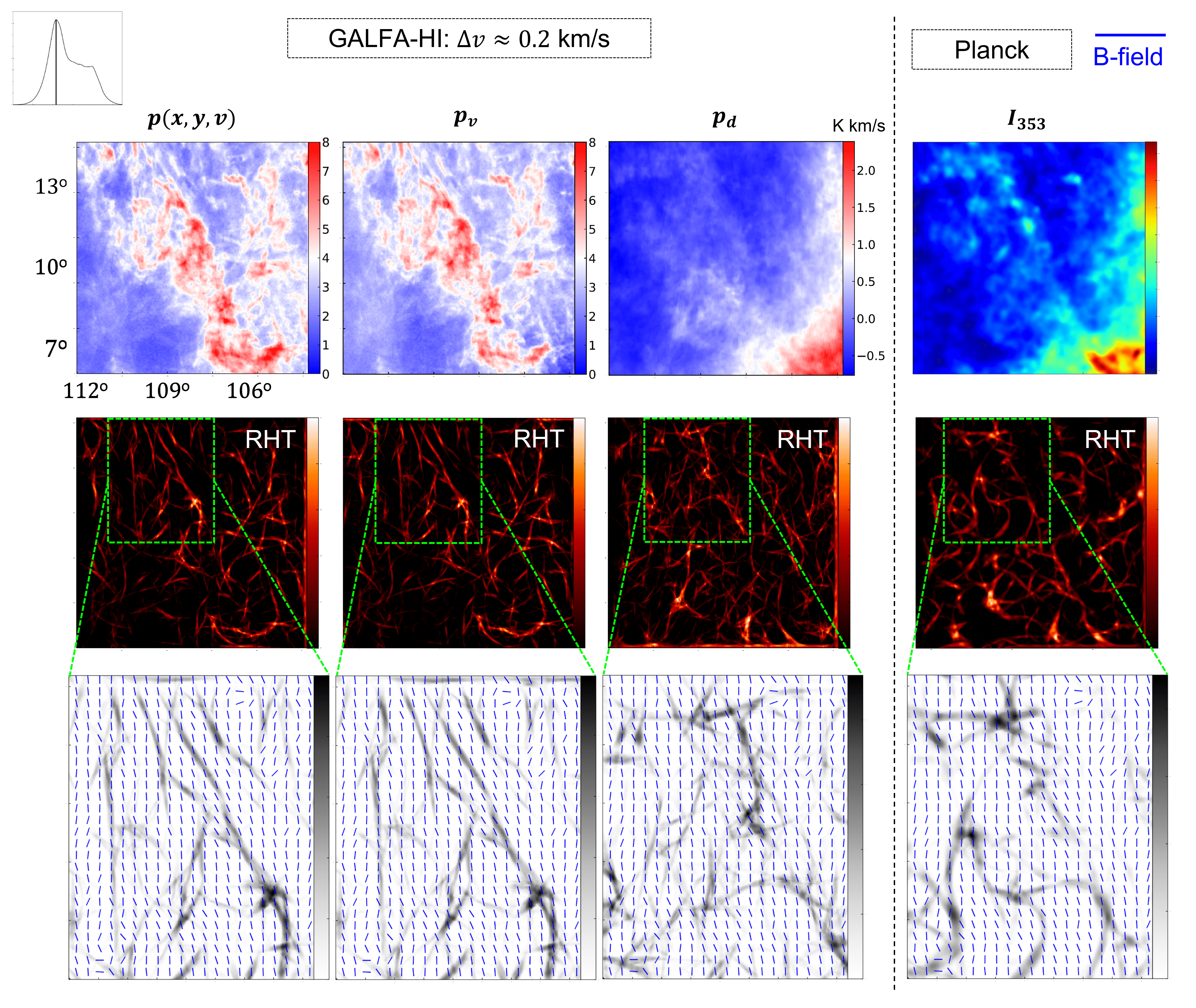}
	\caption{\textbf{Top}: A visual comparison of the thin channel $p$, density contribution $p_d$, velocity contribution $p_v$ from GALFA, and Planck 353 GHz FIR intensity map $I_{353}$. \textbf{Middle}: the corresponding RHT-processed maps. \textbf{Bottom}: a zoom-in region with RHT-identified fibers overlaid with magnetic field orientation (blue segment) inferred from the Planck 353 GHz polarization. The RHT-identified fibers observed in thin channel maps correspond to velocity caustics represented by $p_v$, while the density fibers from $I_{353}$ correspond to $p_d$ image and are mostly aligned perpendicular to the magnetic field.}
	\label{fig:galfa}
\end{figure*}

A direct observational test of whether RHT-fibers originate from velocity caustics or cold neutral filaments is presented in Fig.~\ref{fig:galfa}. We conducted RHT analyses on the raw channel $p$, VDA-decomposed velocity contribution map $p_v$, density contribution map $p_d$, and FIR intensity map $I_{353}$. We repeated the RHT analyses using different parameters (see Fig.~\ref{fig:rht} in Appendix~\ref{app:rht}) to provide the optimal alignment of RHT fibers and magnetic field obtained through polarization observations.

Our results indicated that $p_d$ and $I_{353}$ show a similar appearance and RHT maps. In contrast, no similarity exists between the thin channel maps $p$ and $p_d$ (also $I_{353}$) concerning RHT maps. The RHT maps of thin channels $p$ are comparable to the RHT maps of velocity caustics $p_v$. The perpendicular alignment of RHT-fibers and magnetic fields inferred from Planck 353 GHz polarization is particularly visible in the cases of $p_d$ and $I_{353}$. Such perpendicular alignment is contributed by the density structures that are perpendicular to the magnetic field (see Fig.~\ref{fig:mp3d}) and is expected for supersonic compressible MHD turbulence \citep{2019ApJ...878..157X,2020MNRAS.492..668B}. This effect is natural, as shock compression of gas mostly occurs along the magnetic field in the presence of a strong magnetic field, resulting in structures perpendicular to magnetic fields.

Fig.~\ref{fig:galfa} shows that the thin channel map $p$ resembles $p_v$, which denotes the velocity contribution in $p$. Their RHT maps are similar and the identified RHT fibers align with magnetic fields. This finding is consistent with the numerical analysis presented in \S~\ref{sec:VDAn}. Further tests were conducted by altering the channel width and modifying the $v_{\rm los}$ of $p$ in Figs.~\ref{fig:rht2} and \ref{fig:rht3}. All the results confirm that RHT fibers in a thin channel map arise from the velocity caustics effect. Thus, the success of the RHT technique in tracing magnetic fields reported in \citep{2014ApJ...789...82C,Clark15} is due to the alignment of velocity caustics with the magnetic field. 

Our study confirms the conclusion from MHD turbulence theory (see \citealt{2019tuma.book.....B}), namely, the anisotropy of velocity statistics better traces magnetic field than the anisotropy of density statistics. Thus, it is advantageous to use VDA to remove the density contributions from the data when using RHT to trace the magnetic field.\footnote{In terms of the orientation of filaments, we confirm the density filaments in HI are not different from the density filaments explored in isothermal gas. Namely, these filaments are not necessarily aligned parallel to the magnetic field \citep{2019ApJ...886...17H,2019ApJ...878..157X,2020MNRAS.492..668B}. }

\section{Comparison with earlier studies}
\label{sec:com}

\subsection{Motivation for the present study: Theory-based versus empirical approaches}
The theory of mapping turbulent gas from PPP to PPV space, as formulated in \cite{LP00}, predicts that the intensity structures in spectroscopic channel maps, i.e., slices of PPV cubes, can originate from velocity fluctuations, known as velocity caustics. When the channel width is thin, the intensity structures are dominated by velocity caustics, resulting in a shallower power spectrum than of thick channels \citep{LP00,2016MNRAS.461.1227K}. The theory's quantitative predictions were numerically tested in several observational data sets, including HI, molecular lines $^{12}$CO, $^{13}$CO, $^{18}$CO, and ion lines [S II], [N II], to study turbulent statistics \citep{2006ApJ...653L.125P,2008ApJS..174..202S,2009SSRv..143..357L,2016MNRAS.463.2864A}. The observed changes in the power spectrum of channel maps were consistent with the theoretical expectations. The spectral slopes were utilized to recover the 3D velocity spectrum through the Velocity Channel Analysis (VCA; \citealt{LP00}). The results generally agreed with those obtained through other independent approaches \citep{1995ApJ...443..209A, 2010ApJ...710..853C}.

In this study, we provide tests of the velocity caustics using multi-phase HI MHD simulations. The role of (hydrodynamic) turbulence is to induce velocity crowding. On the other hand, the magnetic field introduces anisotropy to the structures, resulting in striations preferentially aligned with the magnetic field. This anisotropy was already discussed by \cite{2002ASPC..276..182L,2012ApJ...747....5L,2016MNRAS.461.1227K} and tested in \cite{2015ApJ...814...77E}. The anisotropies caused by velocity caustics in thin channel maps can be studied locally by measuring gradients of intensities in channel maps. This resulted in the Velocity Channel Gradient (VChG) technique to trace magnetic field orientation \citep{LY18a}.

\cite{2014ApJ...789...82C} proposed an empirical way of tracing magnetic fields using the linear structures identified by RHT within velocity channel maps. However, \cite{Clark19} denied the existence of velocity caustics and maintained that the striations identified by RHT as fibers are actual cold-density filaments aligned with the magnetic field. While the arguments raised by \cite{Clark19} were addressed in \cite{Yuen19}, the controversy about the nature of HI striations continued with \cite{2019ApJ...886L..13P,2020arXiv200301454K} in denying the existence of velocity caustics and the very validity of the numerically \citep{2003MNRAS.342..325E,2009ApJ...693.1074C,2009ApJ...707L.153P} and observationally \citep{2002ASPC..276..182L,2012ApJ...747....5L,2016MNRAS.461.1227K} tested \cite{LP00} theory. 

The biggest obstacles to achieving a consensus were (1) the difficulties in separating the velocity and density contributions in HI channel maps and (2) the challenges in simulating realistic multi-phase HI gas to provide a direct comparison with the observations. The first problem was recently addressed in \cite{VDA} by introducing the VDA technique to separate velocity and density separation. The paper's HI data analysis complements the present analysis. It also testified to the importance of velocity caustics compared to density filaments. Later, \cite{Ho21} developed state-of-the-art numerical simulations of magnetized multiphase HI. Based on these developments, we find it appropriate to have an objective discussion of the nature of the striation in HI thin channel maps.

\subsection{USM analysis for HI and FIR observations}
\subsubsection{Un-normalized parameter in USM analysis}
\cite{Clark19} and \cite{2020arXiv200301454K} argued that the striation seen in the Galactic 21 cm channel maps arises exclusively from actual cold HI density filaments. They compared unsharp masked (USM) HI intensity structures within spectroscopic channels to Planck FIR observations; the latter revealed the projected density. The USM analysis highlights the contrast along the structures' edges, and these edges were expected to be small-scale structures \citep{2016A&A...595A..37K}. \cite{Clark19} and \cite{2020arXiv200301454K} appealed to the (overlapping) correlation between the USM structures and the FIR map as evidence for supporting their explanation that the striations observed in 21 cm channel maps are actual cold HI filaments, but fully denying velocity caustics effects. However, the correlation is quantified by a parameter $\Delta I_{857}$ (\citealt{Clark19}; see also Appendix~\ref{app: I857})\footnote{$\Delta I_{857}$ is defined as \citep{Clark19}:
	\begin{equation}
		\begin{aligned}
			\Delta I_{857} = \frac{\langle I_{857}\omega\rangle - \langle I_{857}\rangle\langle \omega\rangle}{\langle \omega \rangle}= \langle I_{857}\rangle\left(\frac{\langle I_{857}\omega\rangle}{\langle I_{857}\rangle\langle \omega \rangle}-1\right),
		\end{aligned}
	\end{equation}
	where $I_{857}$ is the intensity of Planck $857$~GHz dust emission and $\omega$ is the USM-measured intensity fluctuations of a given channel with width $\Delta v$.}. The $\Delta I_{857}$'s value of $\sim0.5-0.7$ reported in \cite{Clark19} (see their Fig.~6) is un-normalized, which means $\Delta I_{857}$'s upper bound is not unity (see Appendix~\ref{app: I857}). After proper normalization was introduced, this correlation was found to be insignificant \citep{Yuen19}. 

Instead, we utilize a simple approach to quantify the importance of velocity contribution ($p_v$) and density contribution ($p_d$) in channel maps ($p$). We rely on the normalized covariance coefficient (NCC). Our analysis indicates that the velocity contribution is more significant in thin channels (see Fig.~\ref{fig:ncc}) where the velocity width ($\Delta v$) is less than the velocity dispersion ($\sigma_v$). Below we also present a numerical experiment to demonstrate why this USM analysis is inappropriate for testing velocity caustics from a different angle.

\subsubsection{USM analysis does not rule out velocity caustics}
\begin{figure*}
	\includegraphics[width=0.6\linewidth]{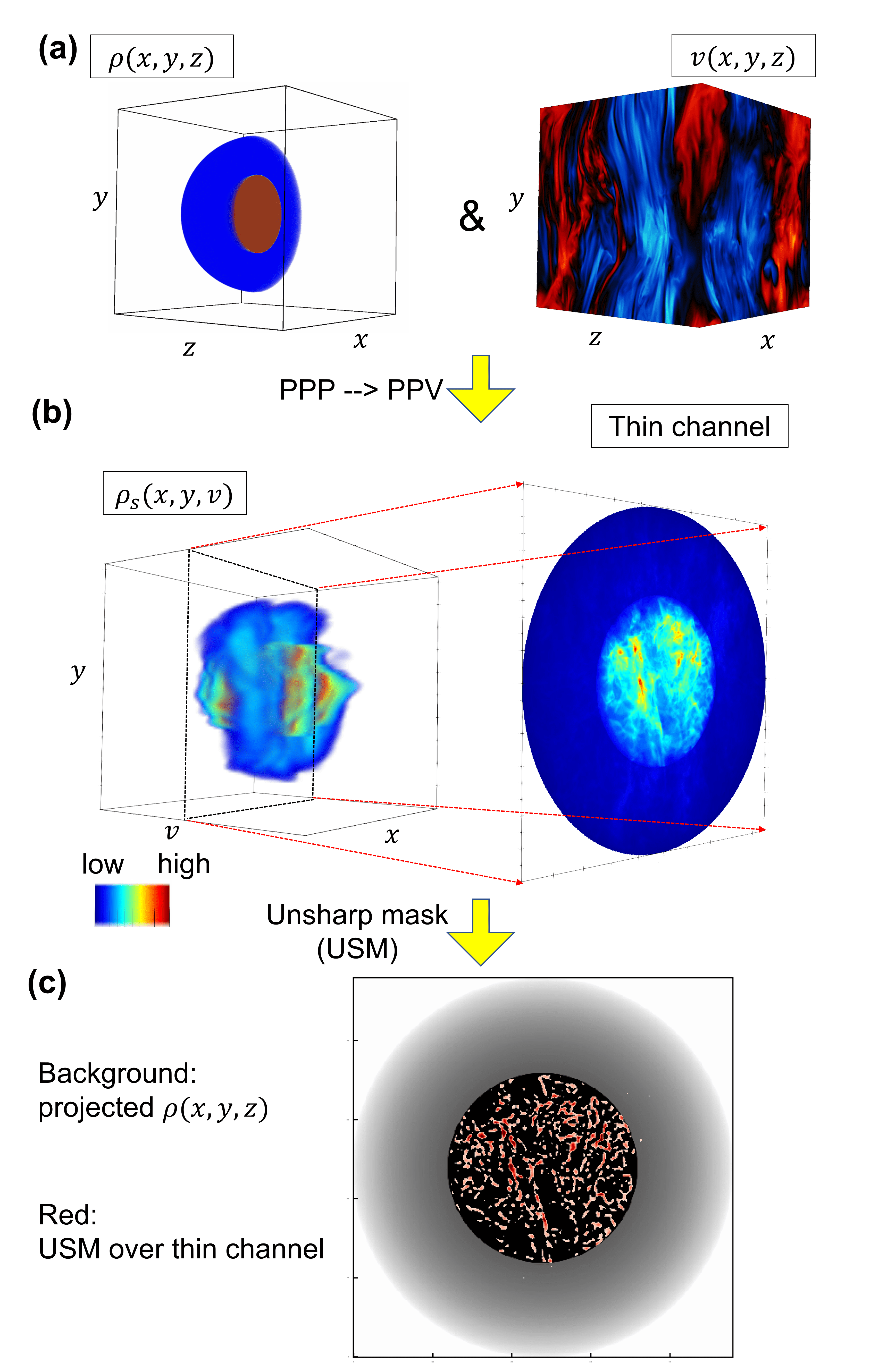}
	\caption{Numerical experiment illustrating the effect of velocity caustics. \textbf{Panel (a)}: 3D visualization of the input density field $\rho(x,y,z)$ and velocity field's LOS component $v(x,y,z)$ in PPV space. The density field $\rho(x,y,z)$ consists of a two-layer hemisphere with a central constant high-density layer and an outskirt constant low-density layer. At the same time, the LOS velocity component $v$ is adopted from the multiphase simulation. The mean magnetic field is oriented along the $y-$direction and thermal broadening is included. \textbf{Panel (b)}: Synthetic PPV cube $\rho(x,y,v)$ generated from $\rho(x,y,z)$ and $v(x,y,z)$. A thin channel slice is extracted from the cube and shown on the right. \textbf{Panel (c)}: Comparison of the projected intensity structures (red) extracted by the USM and the projected real density map (background map: black indicates high density and gray indicates low density). }
	\label{fig:ill}
\end{figure*}

As illustrated in Fig.~\ref{fig:ill}, we generated a synthetic PPV cube from a density field $\rho(x,y,z)$ and velocity field $v(x,y,z)$. The density field $\rho(x,y,z)$ consists of a two-layer hemisphere with a central high-density (ten times higher than the outskirt) layer and an outskirt low-density layer. At the same time, the LOS velocity component $v$ is adopted from the MHD simulation. This two-layer setup aims to eliminate any pre-existing density structures correlated with the magnetic field and examine the effect of high-density structures, which are typically CNM in observation, on the USM analysis. We see the 3D intensity structures in PPV space are highly distorted and do not resemble the real density hemisphere. The 2D intensity structures within a thin channel become filamentary and anisotropically elongated along the magnetic field direction. These filamentary intensity structures are purely created by velocity caustics. We applied the USM to the thin channel in a similar way to \cite{Clark19} and \cite{2020arXiv200301454K}. The USM intensity structures are overlaid with the central high-density area of the projected density hemisphere. 

Instead of denying the existence of velocity caustics, the correlation of overlapping structures supports its presence. In our numerical experiment, if the velocity caustics effect were insignificant, we should only see two large-scale circles in the channel map, and the USM analysis would output only the circular edges. However, as shown in Fig.~\ref{fig:ill}, the USM highlighted many small-scale intensity structures that are purely the results of velocity caustics.

\subsubsection{CNM images are not free from velocity caustics}
In a real scenario, CNM is typically denser than UNM and WNM \citep{1977ApJ...218..148M,1987ARA&A..25..303C,2016A&A...594A.116H,2018ApJS..234....2P}. The USM analysis potentially may separate the dense CNM from multi-phase HI. However, CNM is not free from velocity caustics, unless a CNM cloud can be fully mapped into a channel as an entity. The latter requires that the CNM cloud has a velocity dispersion smaller than the channel width. The median Mach number for CNM was observationally reported as $M_{\rm S}\sim3.7$ \citep{2016A&A...595A..37K}. Assuming the CNM cloud has a typical temperature of 100~K \citep{2001RvMP...73.1031F}, we can estimate the sound speed is $c_s\approx1.17$~km/s and corresponding velocity dispersion $\sigma_v = c_sM_{\rm S}\approx4.33$~km/s, which agrees with the value for a channel can be considered as thin \citep{2001ApJ...551L..53S}, as well as our results in Fig.~\ref{fig:galfa_vda}. When the channel width is smaller than this typical value, the CNM cloud is also affected by the velocity caustics. For instance, as we see in Fig.~\ref{fig:ill}, the high-constant-density hemisphere is strongly distorted after mapping into PPV space. The correlation between USM structures and the projected high-density area arises purely from velocity caustics. This agrees with the picture proposed in \cite{VDA} that while fluctuations in the column density can arise from large scales density fluctuations (corresponding to the central high-density circular area in Fig.~\ref{fig:ill}), small-scale intensity structures are still dominated by velocity caustics (corresponding to the USM-structures in Fig.~\ref{fig:ill}). 

On the other hand, when CNM is mapped into PPV space, it is distributed into different velocity channels due to the velocity caustics effect. This effect does not mean that CNM disappears in the PPV space or that its position on the POS can be different. The correlation always exists when stacking the FIR map with the channels' high-intensity structures (see Fig.~\ref{fig:ill}). Therefore, the observed correlation between the USM-analyzed HI intensity structures within channels and Planck FIR observations cannot distinguish velocity caustics. Similar to the USM analysis, \cite{2019ApJ...886L..13P} and \cite{2020ApJ...899...15M} tried to examine the velocity caustics by comparing the HI intensity structures with the projected location of quasars and column density, respectively. However, for the same reasons mentioned above, these analyses of stacking projected quantity with the HI intensity structures are inappropriate. 

\subsubsection{FIR structures are not always parallel to magnetic field}
\begin{figure}
	\includegraphics[width=1.0\linewidth]{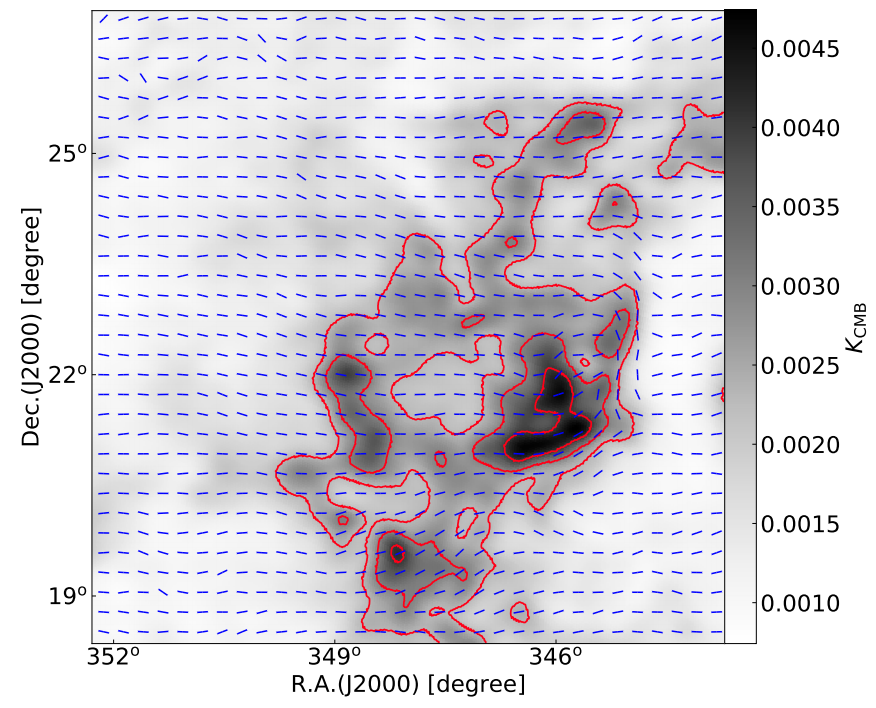}
	\caption{Magnetic field orientation (blue segment) overlaid with the Planck 353 GHz FIR map (background). The red contours start from 0.0023, 0.003, 0.004 $K_{\rm CMB}$.}
	\label{fig:planck}
\end{figure}

Large-scale density structures also tend to follow the magnetic field in globally subsonic conditions \citep{2019ApJ...878..157X}. This very likely happened in the high-latitude $b>30^\circ$ regions selected in earlier studies \citep{Clark19,2019ApJ...886L..13P,2020arXiv200301454K,2020ApJ...899...15M,2023arXiv230316183K}. In Fig.~\ref{fig:planck}, we overlay the magnetic field orientation (inferred from Planck polarization) on the Planck 353 GHz FIR map for a different high-latitude region. Clearly, the FIR density structures are perpendicular to the magnetic field, rather than parallel. These perpendicular FIR structures are ignored in previous studies \citep{Clark19,2020arXiv200301454K}. 

Nevertheless, our analysis was performed for three regions, including high-latitude and low-latitude, to give conclusions (see Figs.~\ref{fig:galfa_vda} and \ref{fig:vda2}). Moreover, we noticed the (i) significance of density contribution can increase in low-latitude HI clouds close to the Galactic disk (see Fig.~\ref{fig:vda2}); (ii) cold HI in our Galaxy concentrates more on the low-latitude ($|b|<30^\circ$) regions (see Fig.~9 in \citealt{2018A&A...619A..58K}). A better alignment of striations and magnetic field should be expected in low-latitude regions if the cold-filament explanation is true \footnote{We would like to note that the use of the USM analysis may pose challenges in low-latitude regions due to the mixing of numerous HI clouds along the LOS. However, as an alternative approach, the magnetic field traced by RHT-fibers should remain unaffected by the mixing, provided that the cold filament explanation is correct. In this scenario, we should expect to observe an improved agreement between the RHT-traced magnetic field and Planck results in low-latitude regions, as well as a positive correlation with channel width. Specifically, a better alignment is expected for the cold filament explanation, when using only thick channels ($\Delta v > \sigma_v$) rather than only thin channels or a mix of thick and thin channels, as such thick channels are certainly dominated by density.}. 

In short, \cite{Clark19,2020arXiv200301454K,2023arXiv230316183K} did not take into account the velocity caustics effect and interpreted those small-scale intensity fluctuations as real cold density filaments. These studies conflict with our numerical experiment and observational evidence which yield the following: (1) the FIR-HI correlation observed through the use of the USM can indeed be generated by velocity caustics (see Fig.~\ref{fig:ill}), while real small-scale density structures can also be present. Thus, the corresponding FIR-HI correlation is not capable to distinguish between structures created by velocity caustics and real density structures within thin channels. (2) The FIR structures do not always align with magnetic fields (see Figs.~\ref{fig:galfa} and \ref{fig:planck}). A good parallel alignment is observed in specific regions \citep{Clark15,Clark19} with $b>30^\circ$, where tangling of the magnetic field along the LOS is not as significant as in the direction towards the Galactic disk.

\subsection{Comparison of numerical simulations}
Some reported results that allegedly contradicted the caustic-based interpretation of the observed striations were obtained using low-resolution simulations \citep{Clark19}. In the numerical experiment by \citet{Clark19}, $128^3$ isothermal simulations were conducted using the RAMSES code \citep{2002A&A...385..337T} for hydrodynamic turbulence, without accounting for the complexity of the actual ISM \citep{1995ApJ...443..209A,2010ApJ...710..853C,Crutcher12,2015A&A...576A.104P,2022ApJ...934....7H}. Further general concerns over the limited inertial range and dynamical range in simulations were raised by \cite{2022arXiv220201610K}, so a convergence study is necessary for this scenario.

To address the limitations, we included high-resolution isothermal MHD simulation with $792^3$ cells, for comparison purposes (see Figs.~\ref{fig:sup04} and Fig.~\ref{fig:sub09}). Our numerical results indicate that velocity caustics are essential and the thermal broadening effect is minimum, especially in the supersonic case, consistent with theoretical expectations \citep{LP00}. Another high-resolution study (480$^3$ to 1200$^3$) in \cite{VDA} supports our findings. 

Furthermore, the multi-phase nature of HI was the main argument in \cite{Clark19} against interpreting striations in terms of velocity caustics, since the theoretical works by \cite{LP00} and others mainly dealt with turbulence in single-phase media. To address this concern, we included non-scale-free multi-phase simulations with 480$^3$ cells in our current study. Our results are consistent with the caustic interpretation of the observed channel map striation. Especially, another multi-phase numerical study in \cite{2023arXiv230504965G} also finds the cold-density filament can be perpendicular to the magnetic field (see Fig.~10 in \citealt{2023arXiv230504965G}). In contrast, no numerical convergence study or multi-phase simulations have been carried out to support the cold-filament interpretation (i.e., cold-filaments always parallel to the magnetic fields) proposed by \cite{Clark19}. 

\section{Discussion}
\label{sec:dis}
\subsection{Origin of HI striations in thin channels}
This study investigates the origin of elongated HI intensity structures within thin channels through a comprehensive set of numerical simulations, including isothermal MHD and multi-phase conditions. These simulations allow us to separate the various physical mechanisms involved. Our isothermal simulations reveal that the velocity caustics effect dominantly creates filamentary intensity structures within thin channels (see Fig.~\ref{fig:sup04}). The velocity caustics also inherit the anisotropy of MHD turbulence, so these structures become aligned parallel to the magnetic fields. The phase transition between CNM, UNM, and WNM has no negative effect on the alignment (see Fig.~\ref{fig:mp}). This observation is also supported by the analysis of GALFA-HI clouds, where we demonstrate that, when the channel width is thin, the HI channel is dominated by velocity contribution (see Fig.~\ref{fig:galfa}).  

We have noticed that the preference for the cold-filament interpretation \citep{Clark19} may stem from a misconception about turbulence. For example, \cite{2022arXiv220201610K} have pointed out that \cite{Clark18} and \cite{2021ApJ...919...53C} have demonstrated that HI striations are coherent structures, but \cite{2022arXiv220201610K} and \cite{2023arXiv230316183K} also claimed that turbulence cannot be invoked to generate coherent structures because they expect turbulence to be a purely random process. However, it is important to note that turbulence is not entirely random. In fact, turbulence is highly structured and can exhibit a wide range of coherent and self-organizing patterns \citep{1983PhLA...99..321T,1988PrAeS..25..231F,1990cp...book.....M}. The reason for this confusion lies in the complex interactions between fluid elements that occur within a turbulent flow. These interactions can produce intricate patterns of vortices, eddies, and waves that evolve in a self-similar and fractal-like manner over a wide range of scales. Our 3D visualization of the velocity field and density field in a turbulent box (see Figs.~\ref{fig:ill} and \ref{fig:mp3d}) clearly shows coherent structures. This was also evident in a number of studies \citep{1981LNP...136....1S,1983PhLA...99..321T,1988PrAeS..25..231F,2009ApJ...693..250B,2013MNRAS.436.1245F,2015PhRvL.115z4503C,2019ApJ...878..157X,2020MNRAS.492..668B,2020ApJ...904..160F,2020NatCo..11.2192K,2023arXiv230315351V}, the largest 10048$^3$-cell turbulence simulation \citep{2021NatAs...5..365F}, and in-situ solar wind measurements of turbulence \citep{2014EGUGA..1610801M,2015EGUGA..1712706P,2016cosp...41E1837S,2017AGUFMSH14A..01K}. One can refer to the literature "Coherent structures in turbulent flows" \citep{1988PrAeS..25..231F}, "Statistical Fluid Mechanics, Volume II: Mechanics of Turbulence" \citep{monin2013statistical}, and recent "Formation and Evolution of Coherent Structures in 3D Strongly Turbulent Magnetized Plasmas" \citep{2023arXiv230315351V}.

It should be noted that the striations in thin channels and real density filaments are different. Our study does not exclude the possibility that real density filaments, especially low-density filaments, can align with magnetic fields in PPP space. We are seeking to answer the origin of HI striations in thin channels and conclude that velocity caustics cannot be ignored. 

\subsection{Density filaments versus spectroscopic striations}
The formation of density filaments is essential in understanding the ISM \citep{2010A&A...518L.100M,2010A&A...518L.102A,2023MNRAS.tmp..495H}. Since Herschel space telescope observations of molecular clouds revealed filamentary patterns of the ISM, several filament-detection algorithms have been proposed \citep{menshchikov2013,juvela2016,alina2022}. These continuum observations contain only density information bringing new insights about, in particular, the star formation process \citep{2022A&A...657L..13P,2022A&A...667L...1A}. The situation, however, is more complicated in HI spectroscopic observations. Instead, velocity caustics raised by the spectroscopic mapping from PPP to PPV space can create virtual intensity striations. These striations do not exist in real space but dominate the intensity structures in thin channels (see Fig.~\ref{fig:vc}). 

The velocity caustics create striations that imprint velocity statistics, while real density filaments contain only density information. Therefore, their properties differ significantly. For velocity statistics, MHD turbulence creates anisotropic velocity structures elongated along magnetic fields \citep{2013SSRv..178..163B,2019tuma.book.....B}, which has been observed in numerical simulations \citep{1981PhFl...24..825M, 1983JPlPh..29..525S,2016JPlPh..82f5301F,2018MNRAS.481.5275T,2020MNRAS.492..668B} and explained by MHD and turbulent reconnection theory \citep{GS95, LV99}. Subsequent numerical studies have confirmed these findings \citep{2000ApJ...539..273C, 2001ApJ...554.1175M, 2003MNRAS.345..325C, 2009ApJ...700...63K}. On the other hand, the behavior of the density field is more complex. Low-density structures can follow velocity statistics \citep{2005ApJ...624L..93B}, while shocks tend to create high-density filaments perpendicular to the magnetic field \citep{2016A&A...586A.135P,2019ApJ...878..157X,2020MNRAS.492..668B}. The thermal instability in HI would increase compressibility and make the formation of perpendicular filaments more vivid. Therefore, when density contribution dominates the thin channels, perpendicular and parallel alignments should be present between the channel's intensity structures and magnetic fields. In contrast, only parallel alignment is expected to be prominent in the case of velocity dominance.

Our 3D visualization of the multi-phase simulation's density and magnetic fields in Fig.~\ref{fig:mp3d} confirms that real high-density filaments can be perpendicular to magnetic fields (see also Fig.~10 in \citealt{2023arXiv230504965G}). This is also observed in the Planck FIR map (see Fig.~\ref{fig:planck}). We identified RHT-fibers for the thin channel maps and found that the perpendicular density filaments seen in PPP space have a marginal contribution to the thin channel RHT-fibers. The RHT-fibers are preferentially aligned along the magnetic field direction, similar to what is observed in GALFA-HI data (see Fig.~\ref{fig:galfa}). This suggests that striations in thin channels mainly arise from velocity caustics and not from density filaments. Therefore, it is important to correctly interpret the results with velocity statistics when using thin channels, particularly in studies aiming to trace Galactic magnetic fields \citep{2023arXiv230204880M,2022ApJ...927...49C,2023arXiv230316183K}. 

\subsection{Implications for HI Related Studies}
\subsubsection{HI striations detected with the Rolling Hough Transform}
We demonstrate that the elongated HI intensity structures within thin channels, and the corresponding RHT-detected fibers, originate from the velocity caustics effect. The properties of the RHT-fibers in thin channels should be dominated by velocity statistics rather than density statistics. On the other hand, the RHT-fibers in thick channels contain more density information. However, density structures are less reliable tracers of magnetic fields because high-density structures can be perpendicular to the magnetic fields (see \citealt{2016A&A...586A.135P,2019MNRAS.485.2825A,2019ApJ...878..157X,2020MNRAS.492..668B} and Fig.~\ref{fig:planck}). 
Therefore, caution should be exercised when using RHT-based polarization studies \citep{2014ApJ...789...82C,Clark15,Clark18,2019ApJ...887..136C} that have completely ignored the velocity caustics effect or have not distinguished between thick ($\Delta v>\sigma_v$) and thin channels ($\Delta v<\sigma_v$). For example, \cite{2023MNRAS.518.4466A} found that the orientation of the filamentary HI structures in the Milky Way toward the Large Magellanic Cloud disagrees with the polarization of foreground stars. Since turbulence and HI are important factors in modeling Galactic foreground polarization, it is essential to correctly interpret the underlying turbulence physics based on either velocity or density (a bad tracer of magnetic fields in HI compared to velocity, due to the presence of perpendicular filaments, see Figs.~\ref{fig:mp3d} and \ref{fig:planck} in this work and Fig.~10 in \citealt{2023arXiv230504965G}). Therefore, physical interpretations that do not consider velocity statistics require careful revision \citep{Clark15,2019ApJ...887..136C,Clark18,2021ApJ...919...53C}.

Furthermore, the outputs of most filament-detection algorithms heavily rely on user-dependent parameters, leading to bias \citep{green2017}. Thus, retrieving the magnetic field's direction from the filament orientation can be uncertain due to parametrization. This parametrization issue can in principle be tackled with a machine learning-based approach (e.g., \citealt{alina2022}). We provide an example of the parameter dependence of the output maps using the RHT method in Appendix~\ref{app:rht}. 

\subsubsection{Magnetic fields traced with Velocity Channel Gradients}
Velocity Channel Gradients (VChGs; \citealt{LY18a}) was developed as a new technique to trace the magnetic fields based on the velocity statistics in thin channels' striations. Its theoretical foundation was questioned by \cite{Clark19}, but this work, as well as other studies \citep{Yuen19,VDA}, has proven the validity of velocity caustics in thin channels. VChGs advantageously holds the promise of obtaining the 3D distribution of the Galactic magnetic fields, including both orientation and strength \citep{2023arXiv230205047H}.  

VChGs is a part of the more general Velocity Gradient Technique (VGT; \citealt{GL17,YL17a,LY18a,HYL18}). VGT can also employ velocities centroids (i.e., moment-1 map or even higher order) to trace the magnetic fields \citep{2020ApJ...905..129H}. Compared to the centroid-based method, VChGs calculated from thin channels are dominated by velocity statistics with less contamination by density, making VChGs a major tool of the VGT. In its turn, VGT is a part of the Gradient Technique (GT) that employs general properties of MHD turbulence for tracing magnetic fields and measuring their strength. The GT has been applied to synchrotron intensity data emissions \citep{2017ApJ...842...30L, 2018ApJ...865...59L, 2020ApJ...901..162H}, Faraday rotation map \citep{2018ApJ...865...59L}, X-ray measurements \citep{2020ApJ...901..162H}, and near-infrared observations \citep{2022MNRAS.511..829H}. The comparison of the results with the available polarization data was successful. The theoretical foundation of all these branches of GT is the anisotropic MHD turbulence \footnote{At low $M_{\rm S}$ the density acts as a passive scalar copying the statistics of velocities \citep{monin2013statistical}.}.

\subsubsection{VChGs and RHT: Comparison and synergy}
RHT is a technique designed to identify filamentary structures in images \citep{2014ApJ...789...82C}, while VChGs is a technique to trace magnetic field direction and obtain their magnetization \citep{LY18a,Lazarian18}. RHT was employed in \cite{Clark15} and subsequent studies to trace magnetic fields using spectroscopic channel maps \citep{Clark18,2019ApJ...887..136C,2021ApJ...919...53C}. They interpreted the results based on density statistics. However, in this paper, we have shown that the fibers observed with RHT mostly arise from velocity caustics.  

VChGs' accuracy in single-phase media has been widely tested using numerical simulations and observations \citep{LY18a,Hu19a,2022A&A...658A..90A,2022MNRAS.511..829H,2022MNRAS.510.4952L,2022arXiv220512084T,2022ApJ...941...92H,2022arXiv221012518S}. The dominance of velocity caustics in cold species with insignificant thermal broadening is robustly true. Although thermal broadening can be significant and phase transition can happen in multi-phase HI, our study with comprehensive numerical and observational tests confirms that the velocity caustics still dominate multi-phase HI thin channels. The theoretical foundation of VGT is justified,  and the corresponding multi-phase numerical test for VGT is given in \cite{2023MNRAS.tmp..209H}.

From the technical point of view, VChGs is a robust technique that does not require adjustable parameters. VChGs' well-understood theoretical foundations, which stem from MHD turbulence theory \citep{GS95,LV99} and spectroscopic mapping theory \citep{LP00}, enable its application to various physical environments. This is achieved by isolating the contributions of density fluctuations, rendering it less susceptible to variations in the orientation of density features \citep{HYL18,LY18a}; removing parasitic contributions of fast MHD modes \citep{2021ApJ...911...53H}; and gauging its accuracy in the presence of non-turbulent (i.e., inflow, outflow, and differential rotation) velocity fields \citep{HYL20,2022MNRAS.511..829H,2022ApJ...941...92H}. 

Our study shows that VChGs compared to RHT, provide higher accuracy of magnetic field testing (see Fig.~\ref{fig:vgt} in Appendix~\ref{app: vgt}). In addition, the distribution of VChGs orientation within an elementary sub-block provides the value of magnetization $M_{\rm A}^{-1}$ (see \citealt{Lazarian18}, and Fig.~\ref{fig:Ma} in Appendix~\ref{app: vgt}), which can then be used to obtain magnetic field strength using the approach in \cite{2020arXiv200207996L,2023arXiv230205047H}. No such capabilities have been reported for the RHT application to channel maps. RHT is one of the ways of identifying filamentary structures that can be successfully employed for its primary purpose once the usage of input parameters is justified (see Appendix~\ref{app:rht}, Fig.~\ref{fig:rht}). 

\subsubsection{VChGs and Machine Learning}
Machine learning (ML) is a rapidly growing field in astrophysics that holds promise for tracing magnetic fields using spectroscopic maps \citep{2023ApJ...942...95X}. However, it is crucial to understand the reasons for the feasibility of such mapping with ML and to be aware of potential pitfalls.

A recent study combined ML learning with velocity gradient in Taurus \citep{2023ApJ...942...95X}, although the authors did not employ the latest modifications of VChGs, making it challenging to make quantitative comparisons. Magnetic field tracing and other areas of observational data analysis can benefit from the ML approach. However, it is vital to comprehend the physical foundation of the applied procedures while utilizing ML. This paper clarifies the connection between striations observed in channel maps and velocity fluctuations aligned by the magnetic field, providing a justification for obtaining magnetic field maps through the ML approach. The space-velocity mapping theory \citep{LP00,2016MNRAS.461.1227K} can help avoid the pitfalls of the brute force ML approach.

The striations observed in thin spectroscopic channels, as shown in this work, are correlated with MHD turbulence \citep{LP00,LY18a}. The ML approach can be further improved and extended by taking into account the theoretical predictions that the VChGs' amplitude \citep{2020ApJ...898...65Y} and curvature \citep{2020ApJ...898...66Y} are correlated with $M_{\rm S}$ and $M_{\rm A}$, respectively. 

\subsubsection{Obtaining Magnetic field strength using HI data}
The Davis–Chandrasekhar–Fermi (DCF) method \citep{1951PhRv...81..890D,1953ApJ...118..113C} is commonly used to estimate the POS magnetic field strength in ISM \citep{HLS21,2021ApJ...907...88P,2021ApJ...913...85H,2021MNRAS.tmp.3119L,2022ApJ...929...27H,2022arXiv220512084T,2023arXiv230316183K}. By assuming magnetic field fluctuations $\delta B$ are fully raised by velocity fluctuations $\sigma_v$ of Alfv\'enic turbulence, the mean POS magnetic field can be approximately obtained from:
\begin{equation}
	B\approx f\frac{\sqrt{4\pi\rho}\sigma_v}{\delta\phi_B},
\end{equation}
here $f\approx0.5$ is a correction factor \citep{2001ApJ...546..980O}, $\rho$ is gas mass density, $\sigma_v$ is velocity dispersion, and $\delta\phi_B$ is the dispersion of magnetic field orientation.
The crucial term in DCF, as well as its modifications \citep{2001ApJ...561..800H,2008ApJ...679..537F,2021A&A...656A.118S,2022MNRAS.514.1575C}, is the ratio of $\sigma_v$ and $\delta\phi_B$. 

For observational implementation, $\sigma_v$ is typically estimated from line width, while $\delta\phi_B$ can be obtained from polarization measurements \citep{2021ApJ...907...88P,HLS21,2021ApJ...913...85H,2021MNRAS.tmp.3119L,2022ApJ...929...27H,2022arXiv220512084T,2022MNRAS.514.1575C}. It is crucial for the correct implementation of DCF that $\delta\phi_B$ invoked in the turbulent dynamo process is on the same scale as $\sigma_v$ \citep{1978mfge.book.....M,2004ApJ...612..276S,2005PhR...417....1B,2016ApJ...833..215X}. This approach to obtaining $\delta\phi_B$ and $\sigma_v$ might overestimate $B$ when the regions under study are only small patches of the cloud \citep{2020arXiv200207996L,2023arXiv230205047H}. The overestimation arises from the fact that (i) $\sigma_v$ estimated from line width corresponds to the velocity fluctuation at injection scale $L_{\rm inj}$ due to the lack of spatial information along the LOS (see Appendix A in \citealt{2023MNRAS.519.3736H}), but (ii) the $\delta \phi_B$ for a small patch with size $l$ corresponds to the fluctuation at scale $l$. Compared to the angle dispersion for the entire cloud (i.e., at turbulence injection scale $L_{\rm inj}$), the dispersion for a small patch is reduced by a factor of $\sim(l/L_{\rm inj})^{1/3}$, assuming Kolmogorov-type turbulence.

Recently, \cite{2023arXiv230316183K} employed the DCF method and its modifications to estimate the POS magnetic field strength in HI striations using thin channels. For this purpose, $\delta\phi_B$ is calculated from the HI-striation-based magnetic field tracing method \citep{2021A&A...654A..91K} and $\sigma_v$ is estimated from HI line width. Firstly, the HI striations are assumed to be real density filaments in \cite{2023arXiv230316183K}, which, however, has been proven not to be the case. Secondly, the $\delta\phi_B$ is calculated at an angular scale of $18'$ (see Sec. 5.1 in \citealt{2023arXiv230316183K}). This corresponds to a length scale of $l\sim0.5$~pc or $l\sim1.3$~pc for their different assumptions of HI LOS distance of 100~pc or 250~pc, respectively. This length scale is much smaller than the typical turbulence injection scale $L_{\rm inj}\sim100$~pc in the ISM \citep{1995ApJ...443..209A, 2010ApJ...710..853C,2022ApJ...934....7H}, leading to a significant overestimation of magnetic field strength. 

On the other hand, \cite{2020arXiv200207996L} suggested the use of sonic and Alfv\'en Mach numbers to obtain the magnetic field strength. This approach mitigated the overestimation in the DCF method \footnote{In addition, the problem with using the smaller patches of data can be mitigated by using the Differential Measure Analysis (DMA) that employs local structure-function for measuring the dispersion \citep{2022ApJ...935...77L,2022arXiv221011023H}.}. A successful application of such an approach to HI is given by \cite{2023arXiv230205047H}. 

\subsubsection{Velocity Channel Analysis}
Velocity Channel Analysis (VCA) was proposed in \cite{LP00} to analyze intensity statistics in PPV spectroscopic data and retrieve the underlying 3D spectra of velocities and densities. The significance of velocity caustics is correlated with the channel map's width, as shown in Fig.~\ref{fig:vc}, and the channel's intensity statistics vary accordingly, which is imprinted in the channel's power spectrum. \cite{LP00} formulated the correlation between the channel's spectrum and velocity and density power spectra, enabling one to determine the spectral index of the velocity and density power spectra through VCA analysis.

Specifically, thin channels' power spectrum is shallower than thick channels due to the intense small-scale striations created by velocity caustics \cite{LP00}. However, \cite{Clark19} challenged this interpretation and questioned the applicability of VCA to multiphase HI, favoring instead that small-scale CNM structures are responsible for the observed HI striations and the corresponding shallow power spectrum. This interpretation conflicts with observational results, as shown by \cite{2020arXiv200301454K}, who decomposed CNM from the Effelsberg Bonn HI Survey (EBHIS) observations and demonstrated that CNM's spectra are steep in the case of thin channels, suggesting small-scale CNMs are rare in thin channels compared to those in thick channels. 

On the other hand, VCA has been successfully applied to both single-phase species and multi-phase HI by different groups (see Tab.~5 in \citealt{2009SSRv..143..357L}, Fig.~1 in \citealt{2006ApJ...653L.125P}, Fig.~18 in \citealt{2008ApJS..174..202S}, and Fig.~1 in \citealt{2016MNRAS.463.2864A}). The numerical and observational analysis presented in this paper further demonstrates that the velocity crowding effect is significant for thin channel maps and is responsible for a significant portion of intensity fluctuations in the channel maps. These results support the applicability of VCA to both single-phase species and multi-phase HI.

\section{Summary}
\label{sec:con}

It is well known that intensity fluctuations observed in PPV can arise from actual density fluctuations and the effect of velocity crowding. 
The latter is the effect that cannot be ignored while analyzing fluctuations in PPV space in general and velocity channel maps in particular. Magnetized turbulence is anisotropic and affects the intensity fluctuations observed in channel maps. The relative role of densities and velocities for the striation observed in thin 21 cm channel maps is explored in this paper. For this purpose, we use synthetic maps of the multi-phase simulations of magnetized HI and compare the results with theoretical expectations and GALFA HI maps.

To better separate the density and velocity contributions, we apply to both synthetic and GALFA HI maps a new statistical tool, Velocity Decomposition Algorithm (VDA), and compare the results obtained with this tool to the input velocity channel maps available. Our major findings are as follows:
\begin{enumerate}
	\item Our analysis of both synthetic maps obtained with multi-phase HI simulations and GALFA HI observations testifies that it is incorrect to disregard the contribution from velocity caustics to the intensity fluctuations in thin channel maps. In fact, we demonstrate that the velocity caustics dominate the striations observed in thin 21 cm velocity channel maps.  
	\item Our study demonstrates the ability of VDA to distinguish between the contribution from velocity caustics and density filaments to the striation observed in thin channel maps obtained with multi-phase HI simulations.
	\item We demonstrate that the VDA-decomposed velocity maps strongly correlate with the original thin channel maps and do not correlate with the column density maps representing density filaments. Applying the VDA analysis to a GALFA-HI cloud reveals that the HI thin channel maps with $\Delta v\approx0.2$ and $4$~km/s are primarily dominated by prominent velocity caustics.
	\item We show that the arguments supporting interpreting striation observed in thin channel maps as arising from the projection of density filaments to the PPV space are not tenable. For instance, the correlation between unsharp-masked HI structures and far infrared emission is not an argument to disregard the velocity caustics effect. 
	\item We confirm that the fibers detected by filament-detection algorithms, such as the RHT, within HI thin channels arise from velocity caustics rather than real density structures. The studies that derive magnetic field orientation and strength employing intensity filaments observed in thin channel maps should be revised to account for the velocity origin of the structures. 
	\item We show that real HI and FIR density structures can align both parallel or perpendicular with magnetic fields, which correspond well to the numerical simulations. On the contrary, the striation arising from velocity caustics is mostly aligned parallel to the magnetic field. The latter corresponds both to the expectations based on theory and numerical simulations. This makes the striation a reliable way to study magnetic fields. 
	\item Our study provides additional evidence that in combination with VDA, VCA can be applied to study turbulence spectra in HI. We also prove that VChGs can successfully trace magnetic fields in HI. 
\end{enumerate}

\section*{Acknowledgements}
We thank Ka Ho Yuen and Blakesley Burkhart for the helpful discussions. Y.H. and A.L. acknowledge the support of NASA ATP AAH7546 and ALMA SOSPADA-016. Financial support for this work was provided by NASA through award 09\_0231 issued by the Universities Space Research Association, Inc. (USRA). This work used SDSC Expanse CPU at SDSC through allocation PHY230032 from the Advanced Cyberinfrastructure Coordination Ecosystem: Services \& Support (ACCESS) program, which is supported by National Science Foundation grants \#2138259, \#2138286, \#2138307, \#2137603, and \#2138296. DA acknowledges the Nazarbayev University Faculty Development Competitive Research Grant Program \#11022021FD2912. D.P. thanks KITP, University of Santa-Barbara for hospitality during COSMICWEB23 program and acknowledges that this research was supported in part by the National Science Foundation under Grant No. NSF PHY-1748958.

%%%%%%%%%%%%%%%%%%%%%%%%%%%%%%%%%%%%%%%%%%%%%%%%%%
\section*{Data Availability}
The data underlying this article will be shared on reasonable request to the corresponding author.
%%%%%%%%%%%%%%%%%%%% REFERENCES %%%%%%%%%%%%%%%%%%

% The best way to enter references is to use BibTeX:

\bibliographystyle{mnras}
\bibliography{example} % if your bibtex file is called example.bib
%\newpage
\appendix

\section{What does the correlation between thin channel maps and FIR map tell us?}
\label{app: I857}
The $\Delta I_{857}$ parameter is introduced by \cite{Clark19} to quantify the correlation between HI thin channel maps and Planck $857$~GHz FIR map. $\Delta I_{857}$ for a given channel with width $\Delta v$ can be written as:
\begin{equation}
	\begin{aligned}
		\Delta I_{857} &= \frac{\langle I_{857}\omega\rangle - \langle I_{857}\rangle\langle \omega\rangle}{\langle \omega \rangle}\\
		&= \langle I_{857}\rangle\left(\frac{\langle I_{857}\omega\rangle}{\langle I_{857}\rangle\langle \omega \rangle}-1\right),
		\label{eq:DeltaI857}
	\end{aligned}
\end{equation}
where $I_{857}$ is the intensity of Planck $857$~GHz dust emission, which is dominated by thermal dust and is proportional to dust column density. $\omega$ is the USM-measured intensity fluctuations of the HI channel. $\Delta I_{857}$ essentially measures the cross-correlation between $I_{857}$ and channel fluctuations. $\Delta I_{857}$ is normalized by the mean USM intensity $\langle \omega \rangle$. However, another normalization by the mean $I_{857}$ is missing, which adds difficulty in making conclusions based on the parameter magnitude.

Below we analytically make the physical meaning of $\Delta I_{857}$ clear. Consider two signals, one, $I_{857}$, proportional to the local column density, and another, $\omega$, proportional to the PPV intensity $\rho_s$. The two can be expressed as:
\begin{equation}
	\begin{aligned}
		I_{857}(\mathbf{X}_1) &\propto \int \rho_s(\mathbf{X}_1,v)dv 
		\equiv \rho_c(\mathbf{X}_1),\\
		\omega(\mathbf{X}_2) &\propto \int^{\Delta v}\rho_s(\mathbf{X}_2,v)dv \equiv \rho_s^{\Delta v}(\mathbf{X}_2,v),
	\end{aligned}
\end{equation}
where $\mathbf{X}=(x,y)$ represents the position vector on the POS. $\rho_c$ is column density and $\rho_s^{\Delta v}(\mathbf{X},v)$ is a PPV channel with width $\Delta v$. Note there $\rho_s(\mathbf{X},v) = \int \rho(\mathbf{X},z) \phi(v-v_{\rm tur}(\mathbf{X},z))dz$ is mapped from density $\rho(\mathbf{X},z)$ in PPP space, where $\phi$ is a Maxwellian distribution of the residual thermal velocity by subtracting turbulent velocity $v_{\rm tur}(\mathbf{X},z)$ from LOS velocity $v$. 

It is clear now that if the turbulent velocities $v_{\rm tur}(\mathbf{X},z)$ are uncorrelated \footnote{This property is numerically and observationally demonstrated (see Fig.~\ref{fig:ncc} and \citealt{VDA,2022arXiv220201610K,2022arXiv220207871Y}).} with the density  $\rho(\mathbf{X},z)$, velocity factor gives rise to the same multiplier $\phi_v^{\Delta v} \equiv \int^{\Delta v} dv\left\langle\phi(v-v_{\rm tur}(\mathbf{X},z))\right\rangle$ in all the averages of Equation~(\ref{eq:DeltaI857}). Then we have:
\begin{equation}
	\begin{aligned}
		\langle\omega\rangle&\propto\left\langle \rho_s^{\Delta v}(v)\right\rangle \approx \phi_v^{\Delta v} \left\langle \rho_c \right\rangle, \\
		\langle I_{857}\omega\rangle&\propto\left\langle \rho_c(\mathbf{X_1}) \rho_s^{\Delta v}(\mathbf{X}_2,v) \right\rangle \approx
		\phi_v^{\Delta v}
		\left\langle \rho_c(\mathbf{X_1}) \rho_c(\mathbf{X_2}) \right\rangle.
		\label{eq:5}
	\end{aligned}
\end{equation}
As a result, the correlation can be written as:
\begin{equation}
	\frac{\langle I_{857}\omega\rangle}{\langle I_{857}\rangle\langle \omega \rangle}\approx  \frac{\left\langle \rho_c(\mathbf{X_1}) \rho_c(\mathbf{X_2}) \right\rangle}
	{\langle \rho_c \rangle^2},
\end{equation}
and accordingly $\Delta I_{857}$ is:
\begin{equation}
	\Delta I_{857} = \langle I_{857}\rangle\left(\frac{\left\langle \rho_c(\mathbf{X_1}) \rho_c(\mathbf{X_2}) \right\rangle}{\langle \rho_c \rangle \langle \rho_c \rangle}-1 \right).
\end{equation}
Apparently, $\Delta I_{857}$ is insensitive to the uncorrelated velocity contribution but measures the level of fluctuations in column density. The value of $\Delta I_{857}$ is not normalized, rather is bounded by $\langle I_{857}\rangle$. Although our considerations were idealized, they do illustrate that the correlation measure in Eq.~\ref{eq:DeltaI857} is not suitable to draw conclusions about the level of velocity caustics in channel maps. 

\section{Isothermal MHD turbulence simulations}
\label{app:mhd}
\begin{figure*}
	\includegraphics[width=1.0\linewidth]{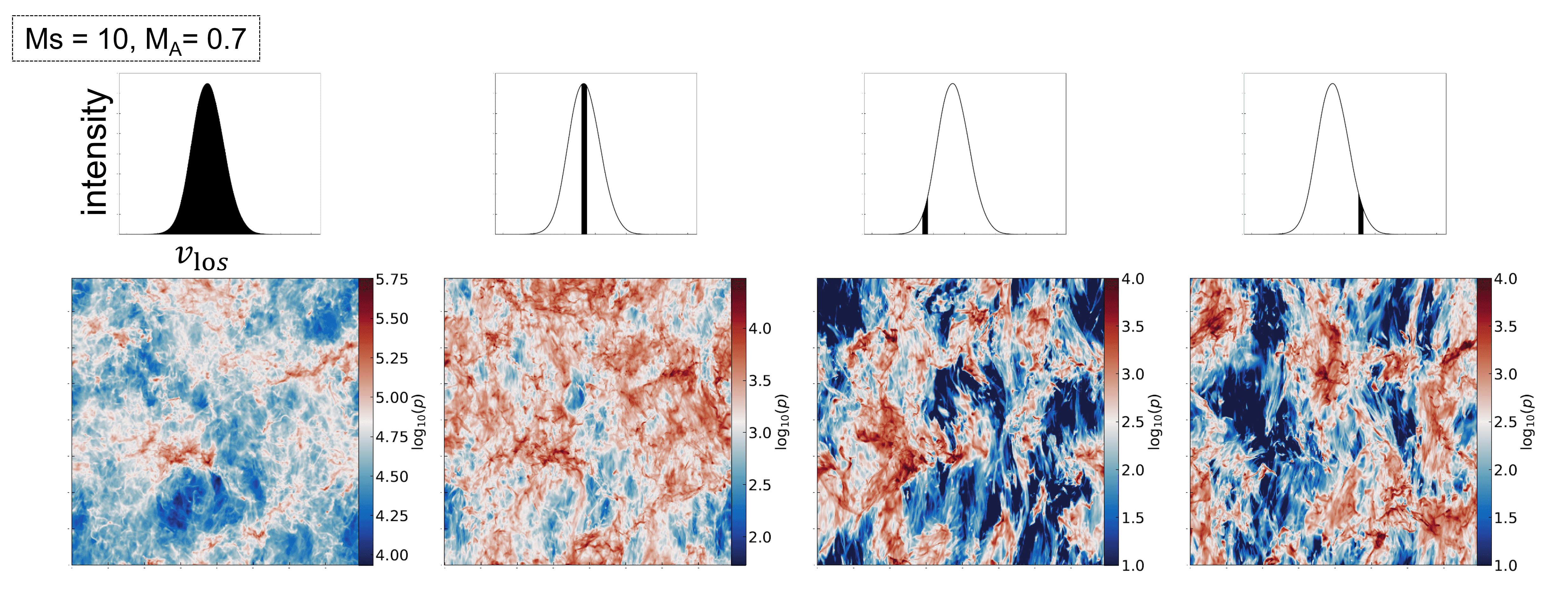}
	\caption{Comparisons of thick velocity channel (left) and thin center and wing channels. The synthetic line is generated by using a non-constant 3D density field $\rho$ from the supersonic MHD simulation. The velocity range used for integration is indicated by the shaded region in the top spectrum. The mean magnetic field is oriented along the vertical $y-$direction and thermal broadening is included.}
	\label{fig:sup04}
\end{figure*}

\begin{figure*}
	\includegraphics[width=1.0\linewidth]{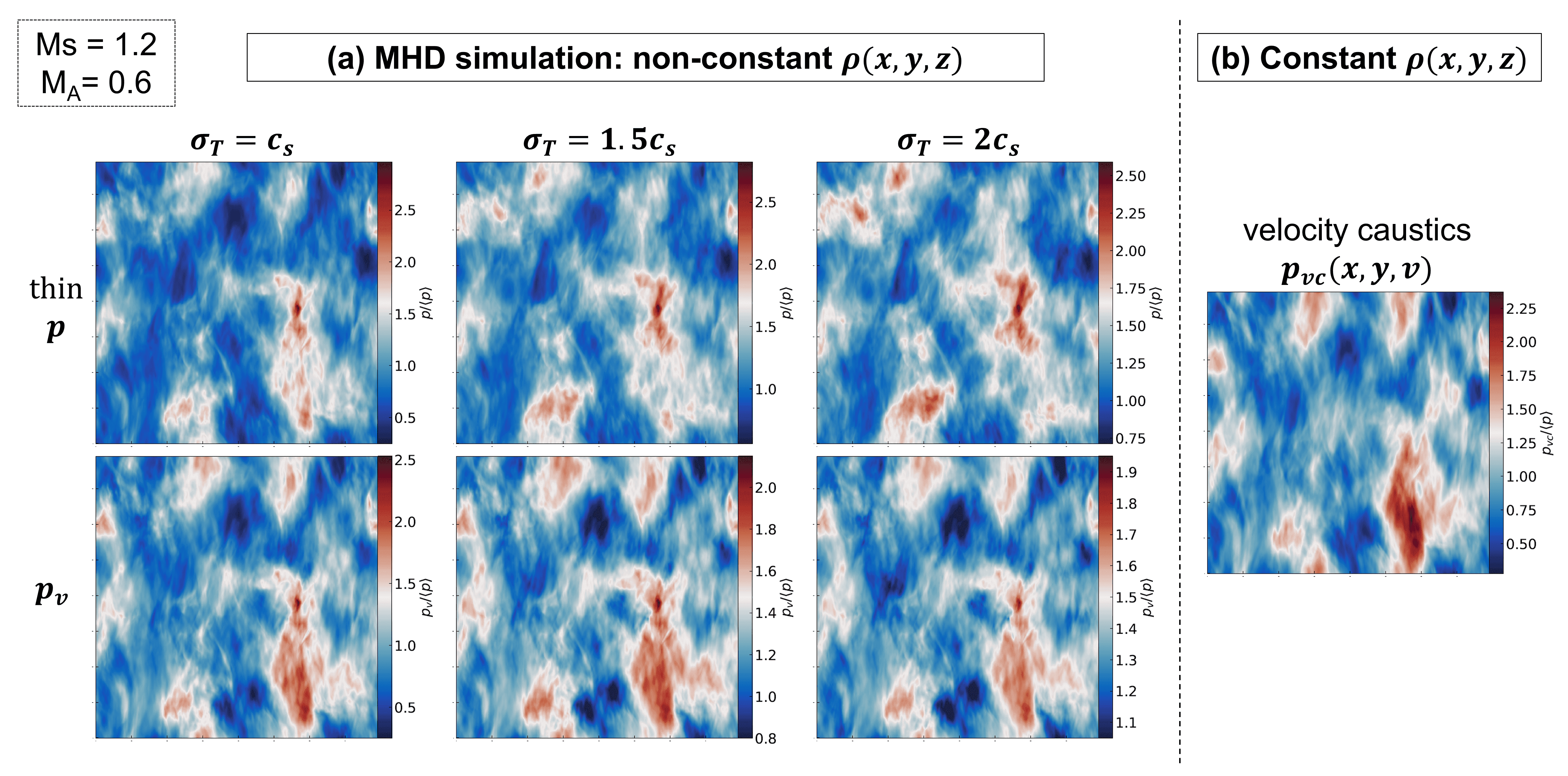}
	\caption{\textbf{Panel (a)}: Comparisons of thin channels $p$ (top) and VDA-decomposed velocity contribution $p_{v}$ with different thermal kernels $\sigma_T$. Larger $\sigma_T$ means stronger thermal broadening. The synthetic line is generated by using a non-constant 3D density field $\rho$ from the $M_{\rm S}=1.2, M_{\rm A}=0.6$ MHD simulation. The channel width is identical to the thin channel used in \ref{fig:sup04}. The mean magnetic field is oriented along the vertical $y-$direction.
		\textbf{Panel (b)}: The thin channel in this panel was generated using a constant density field, thereby eliminating any pre-existing density structures. The structures within the thin channel are solely created by velocity caustics, and thus we refer to it as the velocity caustics channel $p_{vc}$. The integration velocity range is identical to that in panel (a).}
	\label{fig:sub09}
\end{figure*}
We present a supersonic MHD simulation ($M_{\rm S}=10$ and $M_{\rm A}=0.7$) in Fig.~\ref{fig:sup04}, comparing thick and thin (central and wing) channels with thermal broadening included. However, we do not observe any significant resemblance between the thick and thin (central and wing) channels, which differs from the findings of \cite{Clark19} using $128^3$ hydrodynamic simulations. This is expected since, in supersonic conditions (usually in molecular clouds), the thermal speed is considerably smaller than the turbulent velocity, resulting in a minimal contribution of thermal broadening.

The thermal broadening effect is more important for the low-$M_{\rm S}$ case. Thus, we conducted a test to assess the impact of thermal broadening on velocity caustics. To do this, we artificially increased the thermal speed in the isothermal $M_{\rm S}=1.2, M_{\rm A}=0.6$ MHD simulations and compared the wing thin channels $p$ and the pure velocity caustic channel $p_{vc}$, obtained by applying simulation velocities to constant density (as in Fig.~\ref{fig:vc}). In other words, $p_{vc}$ removes any pre-existing density structures. In Fig.~\ref{fig:sub09}, we observe that in the normal $\sigma_T=c_s$ case, the intensity structures in $p$ are similar to those in $p_{vc}$, but differences are also present. When $\sigma_T$ increases (i.e., stronger thermal broadening), the intensity structures in $p$ change further. This is expected (see \S~\ref{sec:vc}), see thermal broadening can decrease the velocity contribution in a channel. However, the results in Fig.~\ref{fig:sub09} suggest that the thermal broadening at a normal level ($\sigma_T=c_s$) is marginal in erasing velocity contribution \footnote{Note that $p_{vc}$ is generated from a constant density cube with $\sigma_T=c_s$. If thermal broadening erases all velocity information, $p_{vc}$ should be only a uniform-intensity map.}.

Nevertheless, VDA was proposed as an effective approach to removing contamination from thermal broadening \citep{VDA}. In Fig.~\ref{fig:sub09}, we test VDA's performance by artificially varying the significance of thermal broadening from $\sigma_T=c_s$ to $\sigma_T=2c_s$. We can see that the VDA-decomposed velocity contribution $p_v$ maps are highly similar in all cases. The $p_v$ maps resemble the $p_{vc}$ map well, even though the thermal broadening effect is stronger than the normal level.

\section{RHT processed thin channel maps}
\label{app:rht}
\begin{figure*}
	\includegraphics[width=1.0\linewidth]{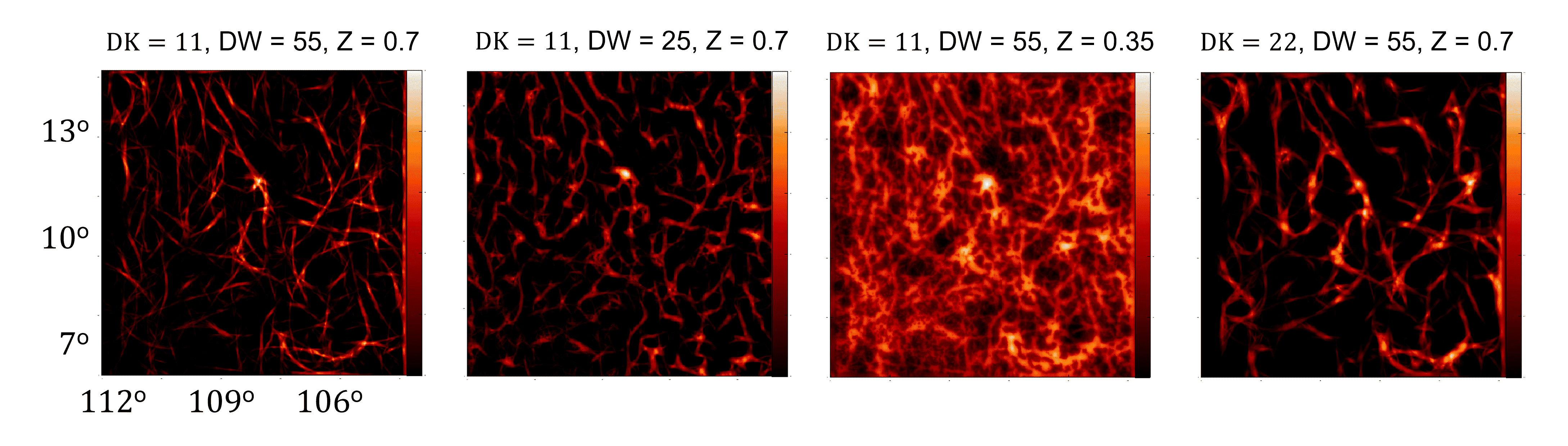}
	\caption{RHT-processed HI thin channel maps (see Fig.~\ref{fig:galfa}) with different input parameters.}
	\label{fig:rht}
\end{figure*}

\begin{figure*}
	\includegraphics[width=1.0\linewidth]{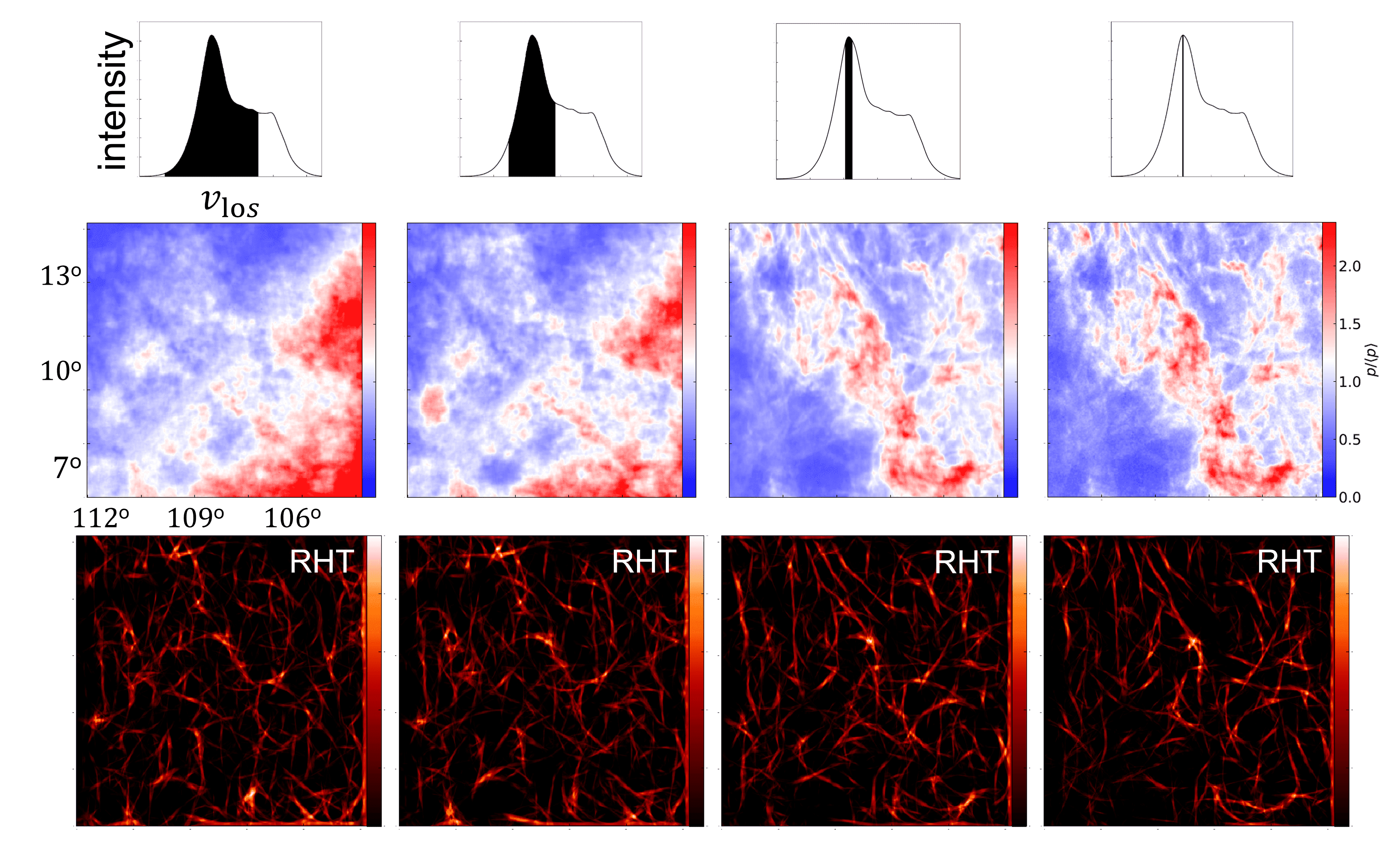}
	\caption{Comparison of HI thin channel maps (middle), as well as their corresponding RHT-maps (bottom) with different channels width ($\Delta v\approx0.2, 4.0, 27, 55$~km/s from right to left). The velocity range used for integration is indicated by the shaded region in the spectrum (top).}
	\label{fig:rht2}
\end{figure*}

\begin{figure*}
	\includegraphics[width=1.0\linewidth]{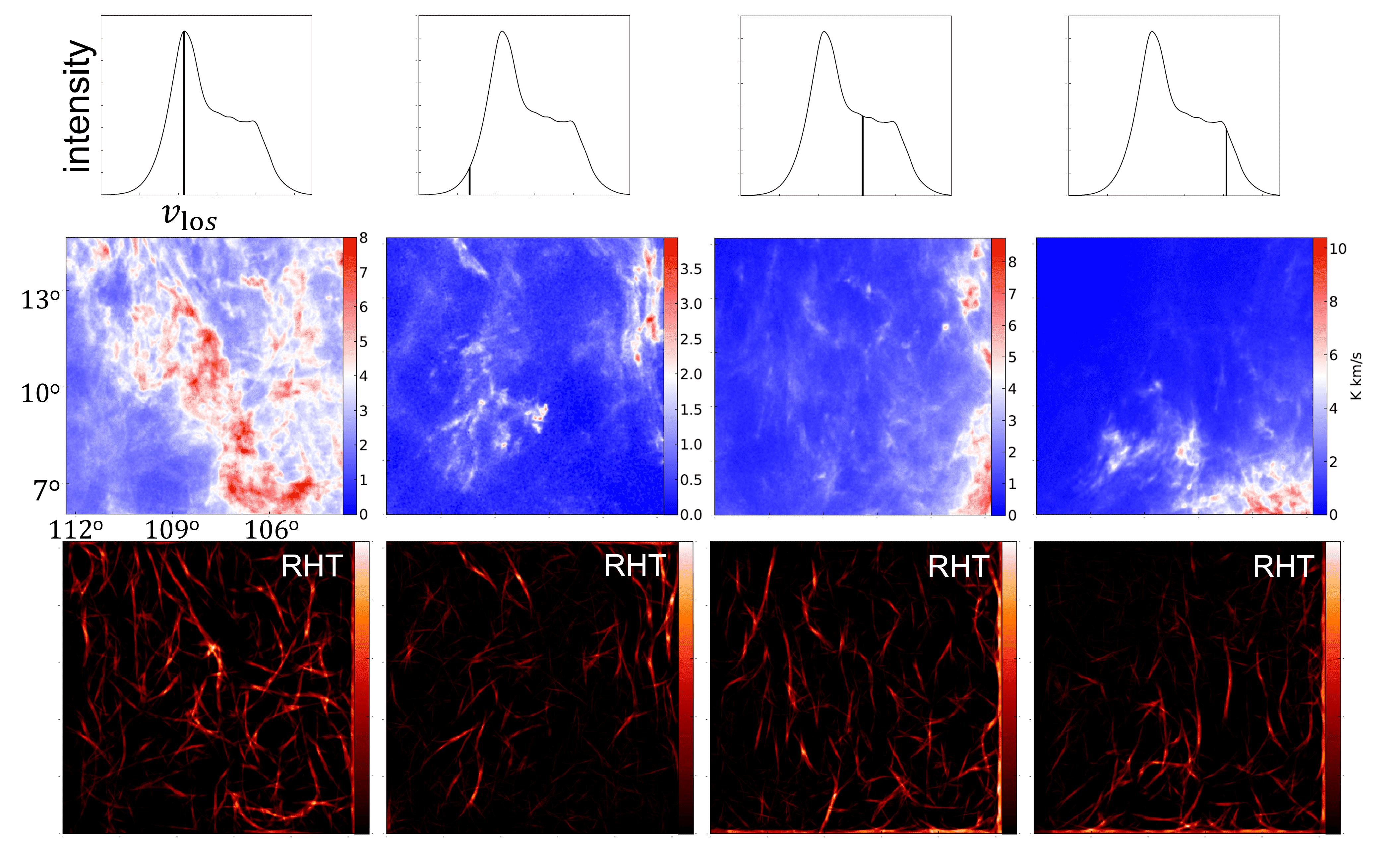}
	\caption{Same as Fig.~\ref{fig:rht2}, but for HI thin channel map $\Delta v\approx0.2$~km/s at different $v_{\rm los}\approx3, -14, 22, 42$~km/s (from right to left).}
	\label{fig:rht3}
\end{figure*}
In order to carry out the RHT analysis, three input parameters are needed: smoothing kernel diameters (DK), a window diameter (DW), and an intensity threshold (Z). Fig.~\ref{fig:rht} displays the RHT-processed thin channel map $p(x,y,z)$, which is used in Fig.~\ref{fig:galfa}, with varying input parameters. Notably, we observed significant differences in the results for the different cases. The parameter DW controls the straightness of the RHT-fibers, while DK determines the thickness. Z regulates the intensity threshold for blanking out low-intensity structures. The RHT-fibers' orientation varies significantly with different input parameters (see Figs.~5, 6, 7 in \citealt{2014ApJ...789...82C}). Especially, the parameter DK would significantly change the aspect ratios of identified RHT-fibers. 

Using the parameters DK = 11, DW = 55, Z = 0.7, we generated RHT-maps for the HI cloud shown in Fig.~\ref{fig:galfa} with different channel widths ($\Delta v\approx0.2, 4.0, 27, 55$km/s). Fig.~\ref{fig:rht2} displays the channel maps and RHT-maps. The channel maps exhibit similar filamentary structures when $\Delta v\approx0.2, 4.0$km/s, while the striations become less noticeable for thicker channels. The RHT-maps ($\Delta v\approx0.2$ and 4.0km/s) are comparable, with RHT-fibers aligning with the magnetic field (as shown in Fig.~\ref{fig:galfa}). The same trend is observed in the other two RHT-maps ($\Delta v\approx27$ and 55km/s), but perpendicular RHT-fibers become more prominent (see Fig.~\ref{fig:galfa}).

It should be noted that the changes in channel maps and RHT-maps could be due to the projection of multiple HI clouds. Specifically, if the cold-filament hypothesis is correct, the perpendicular RHT-fibers should have higher intensities and be visible in other thin channels ($\Delta v\approx0.2$). To investigate this possibility, Fig.~\ref{fig:rht3} presents four thin channels and their RHT-maps (DK = 11, DW = 55, Z = 0.7) located at different $v_{\rm los}\approx3, -14, 23, 43$~km/s. Their LOS velocity difference ($\sim20$km/s) is greater than the typical turbulent velocity dispersion. In all four channels, the HI striations exhibit morphological differences, but all tend to align with the magnetic field (as shown in Fig.~\ref{fig:vgt} for the magnetic field map). The wing channels' intensities ($v_{\rm los}\approx -14, 23, 43$~km/s) are smaller than those in the central channel ($v_{\rm los}\approx3$km/s). In terms of the RHT-maps, no apparent perpendicular RHT-fibers are visible, while such fibers are observed in the upper part of thick channels and the FIR map (as illustrated in Figs.~\ref{fig:galfa} and \ref{fig:rht2}). This simple test excludes the possibility of the LOS projection effect.

\section{VDA decomposed GALFA-HI thin channel}
\label{app: vda}
\begin{figure*}
	\includegraphics[width=1.0\linewidth]{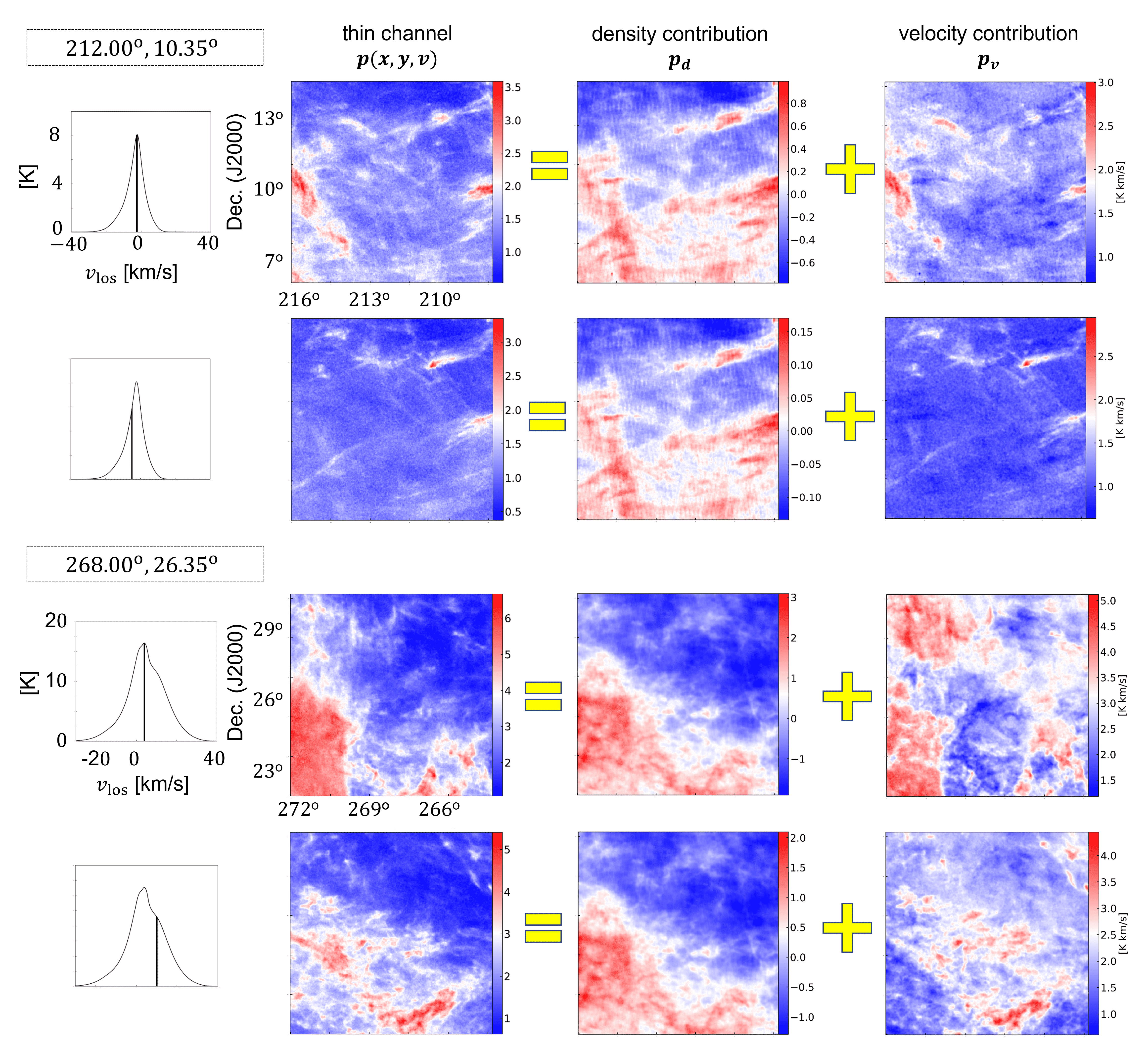}
	\caption{Same as Fig.~\ref{fig:galfa_vda}, but for other two HI clouds with central coordinates are (R.A., Dec.) =  ($212.00^\circ$, $10.35^\circ$) and ($268.00^\circ$, $26.35^\circ$), as well as central $v_{\rm los}\approx-2.3$ and $-10$~km/s for the former and $v_{\rm los}\approx4.0$ and $10$~km/s for the later. The channel width is $\Delta v\approx0.2$~km/s.}
	\label{fig:vda2}
\end{figure*}

Fig.~\ref{fig:vda2} displays a visual comparison between the VDA-separated velocity contribution $p_v$, density contribution $p_d$, and the raw thin channel $p$ for two more HI clouds (one high-latitude and one low-latitude) with central coordinates (R.A., Dec.) =  ($212.00^\circ$, $10.35^\circ$) and ($268.00^\circ$, $26.35^\circ$). The channel width is $\Delta v\approx0.2$~km/s. For the ($212.00^\circ$, $10.35^\circ$) cloud, $p$ and $p_v$ are highly similar with NCC values of $\approx0.63$ for $v_{\rm los}\approx-2.3$~km/s and $\approx0.71$ for $v_{\rm los}\approx-10$~km/s suggesting the dominance of velocity contribution in thin channels. The situation is more complicated for the ($268.00^\circ$, $26.35^\circ$) cloud, the similarity of $p_v$ and $p$ is observed only in the bottom half map for $v_{\rm los}\approx4.0$. This indicates possible equipartition of velocity and density contributions in this particular velocity channel \citep{VDA}. We obtain NCC$\approx0.50$ for this channel, while $\approx0.53$ for the one with $v_{\rm los}\approx10.0$~km/s. 

\section{Comparisons of VChGs and RHT}
\label{app: vgt}
	
\begin{figure*}
		\includegraphics[width=1.0\linewidth]{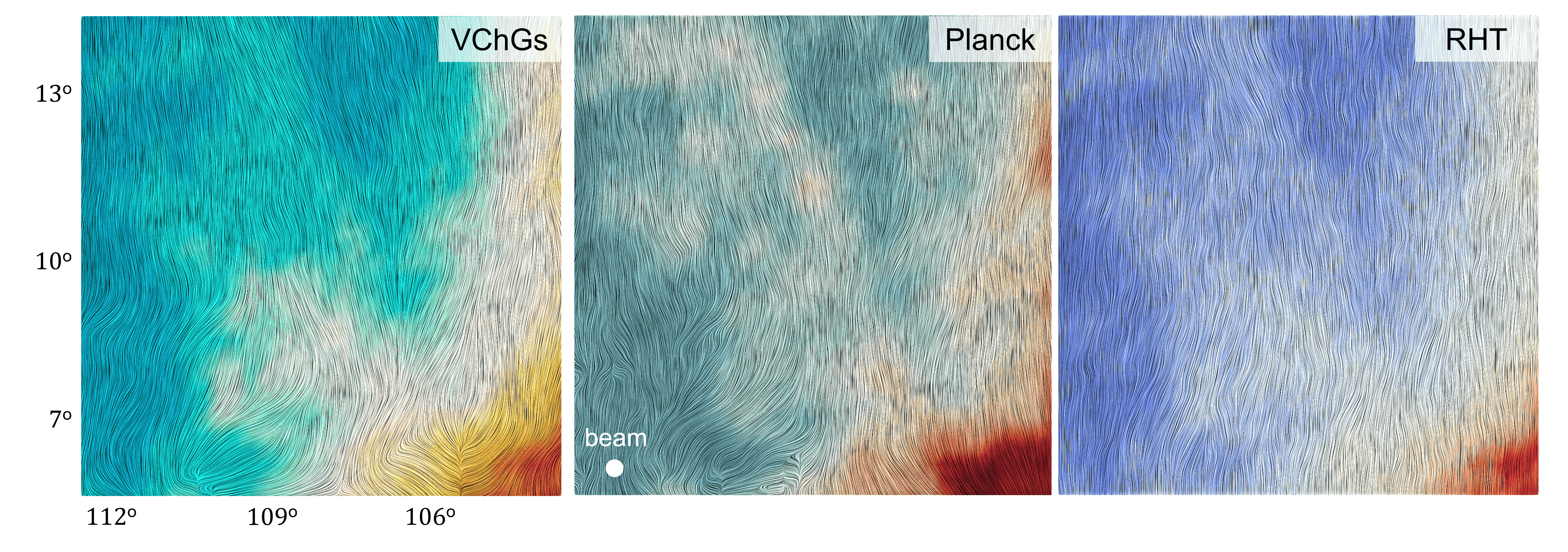}
		\caption{Magnetic field morphology derived from VChGs (left, background image: integrated HI intensity map), Planck 353 GHz polarization (the $2^{\rm nd}$ panel, background image: Planck FIR map), and RHT (the $3^{\rm rd}$ panel, background image: integrated HI intensity map) for the GALFA-HI cloud presented in Fig.~\ref{fig:galfa}.}
		\label{fig:vgt}
\end{figure*}
	
\begin{figure*}
		\includegraphics[width=0.45\linewidth]{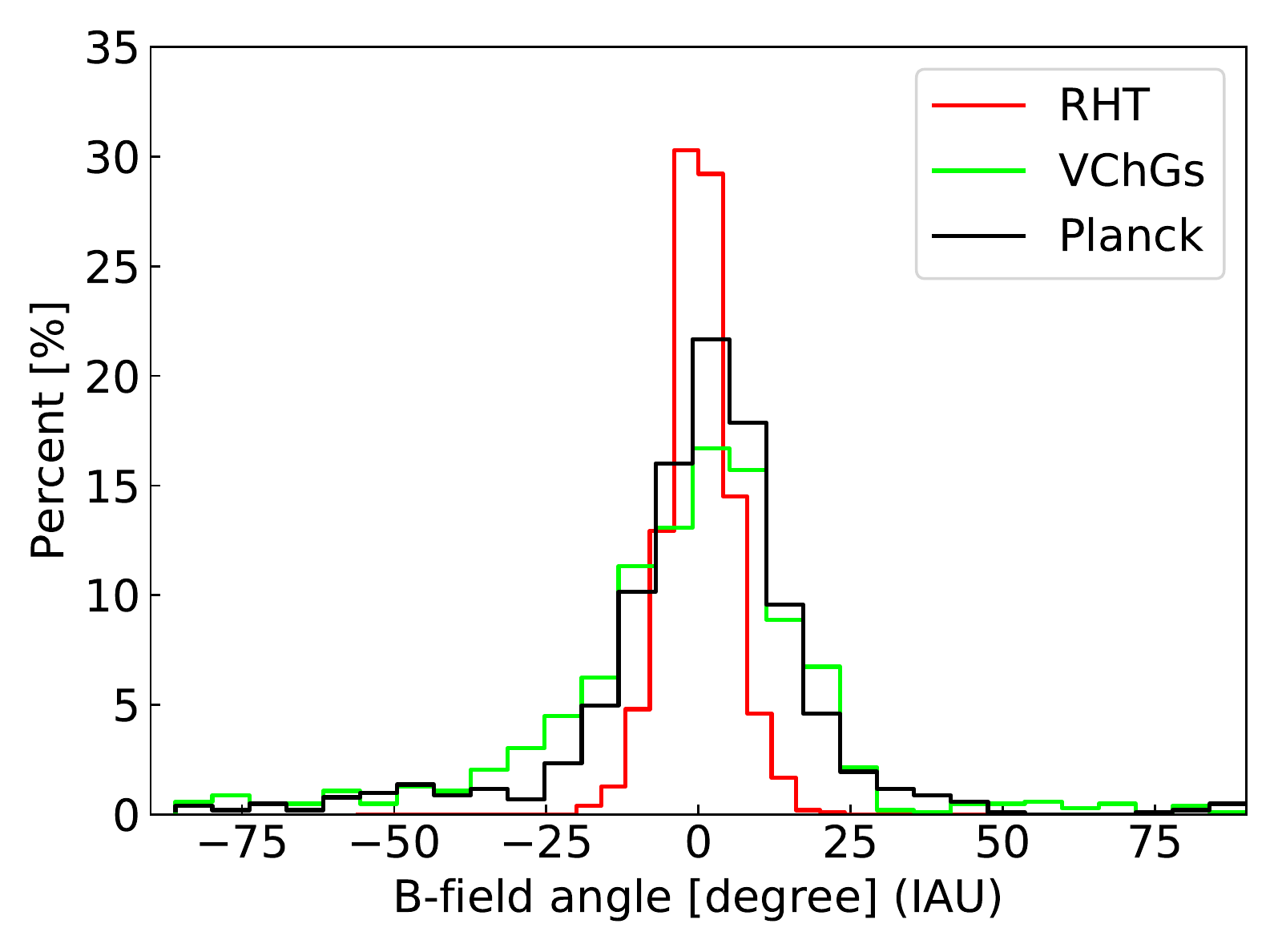}
		\caption{Histogram of magnetic field angle in IAU convention inferred from VChGs, Planck 353 GHz polarization, and RHT for the GALFA-HI cloud presented in Fig.~\ref{fig:vgt}.}
		\label{fig:hist}
\end{figure*}
	
\begin{figure*}
		\includegraphics[width=0.45\linewidth]{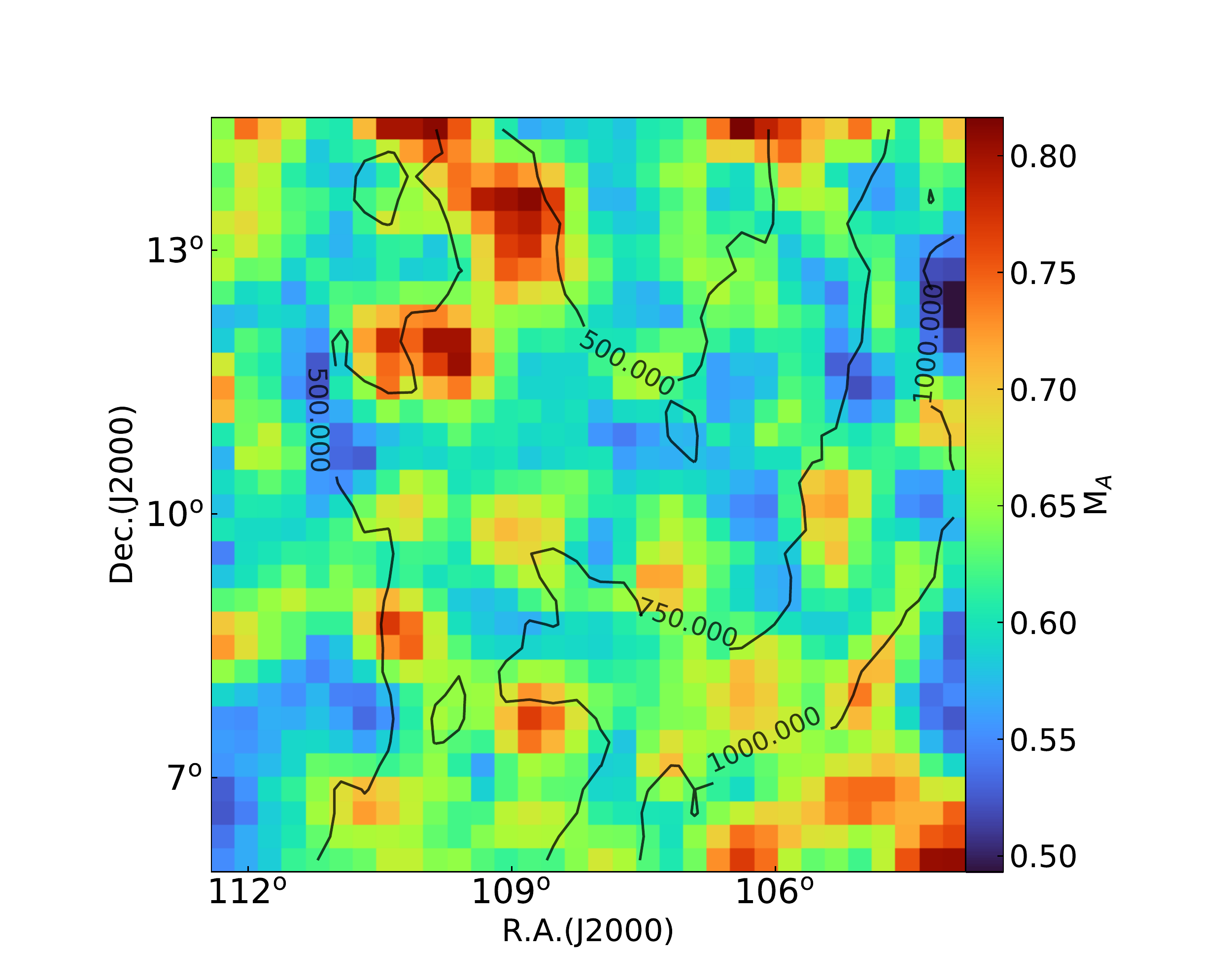}
		\caption{Map of the Alfv\'en Mach number $M_{\rm A}$ distribution derived from VChGs. The contours (in unit of K km/s) highlight prominent structures in the GALFA-HI cloud presented in Fig.~\ref{fig:vgt}.}
		\label{fig:Ma}
\end{figure*}
	
Fig.~\ref{fig:vgt} presents a comparison of magnetic field morphology derived from VChGs, Planck 353 GHz polarization, and RHT. For VChGs calculation, we adopt the recipe used in \cite{2023arXiv230205047H}. We calculate the gradient in every $\Delta v\approx0.2$~km/s channel (gradients are blanked out if their associated HI intensity is less than three times the noise level) and average the gradient over $16\times16$ cells sub-block to get the VChGs orientation. In order fairly compare with Planck polarization, VChGs is then used to construct pseudo-Stokes parameters $Q_{\rm g}$ and $U_{\rm g}$ and integrated over the full velocity range from -40 km/s to 70 km/s. The final POS magnetic field orientation is derived from $\phi_B=\frac{1}{2}\arctan(U_{\rm g},Q_{\rm g})+\frac{\pi}{2}$. 
	
For RHT, we follow \cite{2019ApJ...887..136C} for calculation. We applied a kernel bar (DW) of the length of $25$ pixels and the histogram fraction threshold (Z) was taken at 0.7 of the maximum. The smoothing window (DK) was set to $5$ pixels. The procedure was repeated for every velocity channel at the native GALFA spectral resolution ($\Delta v\approx0.2$~km/s). The RHT angle then is averaged over the velocity range of [-40, 70] km/s and projected into the POS. Finally, we smooth the POS magnetic field orientation inferred from RHT to match the resolution of VChGs. 
	
As shown in Fig.~\ref{fig:vgt}, the magnetic field for this cloud (see Fig.~\ref{fig:galfa}) orients almost along the north-south direction. An apparent variation of magnetic field appears at the lower left corner in Planck and VChGs' maps, while such variation is not observed in the RHT map. In Fig.~\ref{fig:hist}, we plot the histogram of the magnetic field angle in IAU convention inferred from the three methods. We can see VChGs shows better agreement with Planck than RHT. The VChGs and Planck angles are widely spreading from -90 to 90 degrees, while the RHT angle is less dispersed concentrating in the range of [-25, 25] degrees.
	
We define the \textbf{Alignment Measure} (AM) to quantify the alignment of the magnetic field derived from VChGs (also RHT) and the Planck polarization: ${\rm AM}=2\langle\cos^{2} \theta_{r}-\frac{1}{2}\rangle$, where $\theta_r$ is the relative angle between the POS magnetic field inferred from two methods. A value of ${\rm AM} = 1$ indicates a perfect parallel alignment, while ${\rm AM} = -1$ represents a perpendicular alignment. The AM of VChGs and Planck achieves $\sim0.81$, while it is $\sim0.66$ for RHT and Planck. While it is possible to achieve higher AM values by adjusting the input parameters, the physical meanings of the parameters are not clear. Nevertheless, the higher AM value suggests the parameter-free VChGs is more accurate in tracing magnetic fields. 
	
On the other hand, VChGs is also able to derive the POS Alfv\'en Mach number $M_{\rm A}$ distribution. The physical reason behind it is that a strong magnetic field (i.e., small $M_{\rm A}$) induces strong anisotropy, which corresponds to a narrow histogram (i.e., small dispersion) of the velocity gradient orientation. For a weak magnetic field (i.e., large $M_{\rm A}$), the histogram tends to be isotropic with a large dispersion. This relation of $M_{\rm A}$ and velocity gradient orientation's dispersion was proposed in \cite{Lazarian18} and successfully used to derive the $M_{\rm A}$ distribution in 3D over our Galaxy \citep{2023arXiv230205047H} with the assistance of the Galactic rotational curve. Here we follow the same recipe used in \cite{2023arXiv230205047H} (see their Sec.~3) to find the $M_{\rm A}$ distribution for the GALFA-HI cloud presented in Fig.~\ref{fig:vgt}. As shown in Fig.~\ref{fig:Ma}, the $M_{\rm A}$ is calculated for every $16\times16$ pixels. We find this HI cloud is globally sub-Alfv\'enic with a mean $M_{\rm A}$ of $\sim0.63$.

% Alternatively you could enter them by hand, like this:
% This method is tedious and prone to error if you have lots of references
%\begin{thebibliography}{99}
%\bibitem[\protect\citeauthoryear{Author}{2012}]{Author2012}
%Author A.~N., 2013, Journal of Improbable Astronomy, 1, 1
%\bibitem[\protect\citeauthoryear{Others}{2013}]{Others2013}
%Others S., 2012, Journal of Interesting Stuff, 17, 198
%\end{thebibliography}

%%%%%%%%%%%%%%%%%%%%%%%%%%%%%%%%%%%%%%%%%%%%%%%%%%
		
%%%%%%%%%%%%%%%%% APPENDICES %%%%%%%%%%%%%%%%%%%%%
		
%%%%%%%%%%%%%%%%%%%%%%%%%%%%%%%%%%%%%%%%%%%%%%%%%%

		% Don't change these lines
		\bsp	% typesetting comment
		\label{lastpage}
	\end{document}